\documentstyle[]{article}
\def\hybrid{\topmargin 0pt      \oddsidemargin 0pt
        \headheight 0pt \headsep 0pt
        \textwidth 6.5in        % US paper
        \textheight 9.0in         % US paper
        \marginparwidth 0.0in
        \parskip 5pt plus 1pt   \jot = 1.5ex}
\catcode`\@=11
\def\marginnote#1{}

\newcount\hour
\newcount\minute
\newtoks\amorpm
\hour=\time\divide\hour by60
\minute=\time{\multiply\hour by60 \global\advance\minute by-\hour}
\edef\standardtime{{\ifnum\hour<12 \global\amorpm={am}%
        \else\global\amorpm={pm}\advance\hour by-12 \fi
        \ifnum\hour=0 \hour=12 \fi
        \number\hour:\ifnum\minute<10 0\fi\number\minute\the\amorpm}}
\edef\militarytime{\number\hour:\ifnum\minute<10 0\fi\number\minute}

\def\draftlabel#1{{\@bsphack\if@filesw {\let\thepage\relax
   \xdef\@gtempa{\write\@auxout{\string
      \newlabel{#1}{{\@currentlabel}{\thepage}}}}}\@gtempa
   \if@nobreak \ifvmode\nobreak\fi\fi\fi\@esphack}
        \gdef\@eqnlabel{#1}}
\def\@eqnlabel{}
\def\@vacuum{}
\def\draftmarginnote#1{\marginpar{\raggedright\scriptsize\tt#1}}

\def\draft{\oddsidemargin -.5truein
        \def\@oddfoot{\sl preliminary draft \hfil
        \rm\thepage\hfil\sl\today\quad\militarytime}
        \let\@evenfoot\@oddfoot \overfullrule 3pt
        \let\label=\draftlabel
        \let\marginnote=\draftmarginnote
   \def\@eqnnum{(\theequation)\rlap{\kern\marginparsep\tt\@eqnlabel}%
\global\let\@eqnlabel\@vacuum}  }
\font\teneuf=eufm10  scaled  1200
\font\seveneuf=eufm7 scaled  1200
\font\fiveeuf=eufm5  scaled  1200

\newfam\euffam
\textfont\euffam=\teneuf  \scriptfont\euffam=\seveneuf
  \scriptscriptfont\euffam=\fiveeuf

\def\hexnumber@#1{\ifnum#1<10 \number#1\else
 \ifnum#1=10 A\else\ifnum#1=11 B\else\ifnum#1=12 C\else
 \ifnum#1=13 D\else\ifnum#1=14 E\else\ifnum#1=15 F\fi\fi\fi\fi\fi\fi\fi}

\def\got{\ifmmode\let\next\got@\else
 \def\next{\errmessage{Use \string\got\space only in math mode}}\fi\next}
\def\got@#1{{\got@@{#1}}}
\def\got@@#1{\fam\euffam#1}

\newfont{\lgot}{eufm10 scaled 2000}%

\newfont{\Bbb}{msbm10 scaled 1\@ptsize00}

\newcommand{\ZZ}{\mbox{\Bbb Z}}
\newfont{\Bbbb}{msbm7 scaled 1\@ptsize00}

\font\teneufm=cmmib10
\font\seveneufm=cmmib7
\font\fiveeufm=cmmib5
\def\bfit#1{{\textfont1=\teneufm\scriptfont1=\seveneufm
\scriptscriptfont1=\fiveeufm
\mathchoice{\hbox{$\displaystyle#1$}}{\hbox{$\textstyle#1$}}
{\hbox{$\scriptstyle#1$}}{\hbox{$\scriptscriptstyle#1$}}}}
\font\sevenmsa=msam6
\newfam\msafam
\textfont\msafam=\sevenmsa
\def\hexnumber@#1{\ifnum#1<10 \number#1\else
\ifnum#1=10 A\else\ifnum#1=11 B\else\ifnum#1=12 C\else
\ifnum#1=13 D\else\ifnum#1=14 E\else\ifnum#1=15 F\fi\fi\fi\fi\fi\fi\fi}
\def\msa@{\hexnumber@\msafam}
\mathchardef\blacktriangleright="3\msa@49
\mathchardef\blacktriangleleft="3\msa@4A
\newdimen\linethick  \linethick=0.4pt
\newdimen\hboxitspace    \hboxitspace=5pt
\newdimen\vboxitspace    \vboxitspace=5pt

\def\fr#1{%
\beq\new
\vcenter{
\hrule height\linethick
           \hbox{\vrule width\linethick
                 \kern\hboxitspace
                 \vbox{\kern\vboxitspace
                       \hbox{$\begin{array}{c}\displaystyle#1
          \end{array}$}%
                       \kern\vboxitspace}%
                 \kern\hboxitspace
                 \vrule width\linethick}%
           \hrule height\linethick}%
\eeq}
\def\numberbysection{\@addtoreset{equation}{section}
        \def\theequation{\thesection.\arabic{equation}}}
\numberbysection

\newcommand{\sect}[1]{\setcounter{equation}{0}\section{#1}}
\renewcommand{\theequation}{\thesection.\arabic{equation}}
\newcommand{\l@qq}[2]{\addvspace{2em}
 \hbox to\textwidth{\hspace{1em}\bf #1 \dotfill #2}}

\def\appname{Appendix}
\newcounter{app}
\def\theapp{\Alph{app}}
\def\app{\par
   \addvspace{4ex}
   \@afterindentfalse
  \secdef\@app\@dapp}
\def\@app[#1]#2{\ifnum \c@secnumdepth >\m@ne
        \refstepcounter{app}
        \addcontentsline{toc}{app}{\theapp
        \hspace{1em}#1}\else
      \addcontentsline{toc}{app}{ #1}\fi
   {\parindent \z@ \raggedright
    \Large \bf \appname~\theapp .
   \Large  \bf \hspace{1em}    #2}\nobreak
   \vskip 4ex   \noindent
\setcounter{equation}{0}
\def\theequation{\Alph{app}.\arabic{equation}}}
\def\@dapp#1{%
{\parindent \z@ \raggedright  \bf #1}\par\nobreak}
\def\l@app#1#2{\addpenalty{\@secpenalty}%
   \addvspace{1em plus\p@}%
   \begingroup
   \@tempdima 3em
     \parindent \z@ \rightskip \@pnumwidth
     \parfillskip -\@pnumwidth
     { \bf
     \leavevmode
     #1\hfil \hbox to\@pnumwidth{\hss #2}}\par
     \nobreak
   \endgroup}
\newcounter{sapp}[app]
\def\thesapp{\Alph{app}.\arabic{sapp}}
\def\sapp{\par
   \@afterindentfalse
  \secdef\@sapp\@dsapp}
\def\@sapp[#1]#2{\ifnum \c@secnumdepth >\m@ne
        \refstepcounter{sapp}
        \addcontentsline{toc}{sapp}{\thesapp
        \hspace{1em}#1}\else
      \addcontentsline{toc}{sapp}{ #1}\fi
   {\parindent \z@ \raggedright
    \large \bf \thesapp
   \large  \bf \hspace{1em}    #2}\nobreak
   \vskip 4ex   \noindent
\def\theequation{\Alph{app}.\arabic{equation}}}
\def\@dsapp#1{%
{\parindent \z@ \raggedright  \bf #1}\par\nobreak}
\def\l@sapp#1#2{\addpenalty{\@secpenalty}%
   \begingroup
   \@tempdima 3em
     \parindent \z@ \rightskip \@pnumwidth
     \parfillskip -\@pnumwidth
     { \hspace{1em}
     \leavevmode
     #1 \hfil \dotfill \hbox to\@pnumwidth{\hss #2}}\par \nobreak
     \endgroup}
\def\titlepage{\@restonecolfalse\if@twocolumn\@restonecoltrue\onecolumn
     \else \newpage \fi \thispagestyle{empty}\c@page\z@
}

\def\endtitlepage{\if@restonecol\twocolumn \else  \fi
        \def\thefootnote{\arabic{footnote}}
        \setcounter{footnote}{0}}  %\c@footnote\z@ }
\relax

\hybrid

\makeatletter
\newdimen\normalarrayskip              % skip between lines
\newdimen\minarrayskip                 % minimal skip between lines
\normalarrayskip\baselineskip
\minarrayskip\jot
\newif\ifold             \oldtrue            \def\new{\oldfalse}
\def\arraymode{\ifold\relax\else\displaystyle\fi} % mode of array enrties
\def\eqnumphantom{\phantom{(\theequation)}}     % right phantom in eqnarray
\def\@arrayskip{\ifold\baselineskip\z@\lineskip\z@
     \else
     \baselineskip\minarrayskip\lineskip1\baselineskip\fi}

\def\@arrayclassz{\ifcase \@lastchclass \@acolampacol \or
\@ampacol \or \or \or \@addamp \or
   \@acolampacol \or \@firstampfalse \@acol \fi
\edef\@preamble{\@preamble
  \ifcase \@chnum
     \hfil$\relax\arraymode\@sharp$\hfil
     \or $\relax\arraymode\@sharp$\hfil
     \or \hfil$\relax\arraymode\@sharp$\fi}}

\def\@array[#1]#2{\setbox\@arstrutbox=\hbox{\vrule
     height\arraystretch \ht\strutbox
     depth\arraystretch \dp\strutbox
     width\z@}\@mkpream{#2}\edef\@preamble{\halign \noexpand\@halignto
\bgroup \tabskip\z@ \@arstrut \@preamble \tabskip\z@ \cr}%
\let\@startpbox\@@startpbox \let\@endpbox\@@endpbox
  \if #1t\vtop \else \if#1b\vbox \else \vcenter \fi\fi
  \bgroup \let\par\relax
  \let\@sharp##\let\protect\relax
  \@arrayskip\@preamble}
\def\eqnarray{\stepcounter{equation}%
              \let\@currentlabel=\theequation
              \global\@eqnswtrue
              \global\@eqcnt\z@
              \tabskip\@centering
              \let\\=\@eqncr
              $$%
 \halign to \displaywidth\bgroup
    \eqnumphantom\@eqnsel\hskip\@centering
    $\displaystyle \tabskip\z@ {##}$%
    &\global\@eqcnt\@ne \hskip 2\arraycolsep
         $\displaystyle\arraymode{##}$\hfil
    &\global\@eqcnt\tw@ \hskip 2\arraycolsep
         $\displaystyle\tabskip\z@{##}$\hfil
         \tabskip\@centering
    &{##}\tabskip\z@\cr}
\makeatother

\def\bea{\begin{eqnarray}}
\def\eea{\end{eqnarray}}

\def\beq{\begin{equation}}
\def\eeq{\end{equation}}
\def\be{\beq\new\begin{array}{c}}
\def\ee{\end{array}\eeq}
\def\2{{1\over 2}}
\def\stackreb#1#2{\mathrel{\mathop{#2}\limits_{#1}}}

\def\balpha{{\bfit\alpha}}
\def\bbeta{{\bfit\beta}}

\def\d{\partial}

\def\<{\langle}
\def\>{\rangle}
\def\ov{\overline}

\newcommand{\re}[1]{{\ref{#1}}}
\newcommand{\lab}[1]{\label{#1}}
\newcommand{\bi}[1]{\bibitem{#1}}
\newcommand{\ci}[1]{\cite{#1}}

\def\sem{\raise1pt\hbox{$\scriptscriptstyle >\!$}\:\!\!\tl}
\def\dr{\raise1pt\hbox{$\scriptscriptstyle >\!$}\!\!\btl}
\def\c{\epsilon}
\def\ep{\epsilon}
\def\f{1\over }
\def\ov{\overline}

\begin{document}

%\draft                               %SWITCH ON/OFF DRAFT VERSION%

\begin{titlepage}
\setcounter{footnote}0
\begin{center}
\hfill ITEP/TH-1/95\\
\hfill FIAN/TD-19/95\\
%\hfill \today\\
\hfill hep-th/9606144\\
\vspace{0.3in}
{\LARGE\bf Faces of Relativistic Toda Chain}
\\[.4in]
{\Large S.Kharchev\footnote{E-mail
address:kharchev@vxitep.itep.ru}}$\phantom{hj}^{\dag}$, {\Large A.
Mironov\footnote{E-mail address:  mironov@lpi.ac.ru,
mironov@nbivax.nbi.dk}}$\phantom{hj}^{\ddag,\ \dag}$, {\Large A.
Zhedanov\footnote{E-mail address:
zhedanov@host.dipt.donetsk.ua}}$\phantom{hj}^{\sharp}$\\
\bigskip

\bigskip

\begin{quotation}{

$\phantom{hj}^{\dag}$ --
{\it ITEP, Bol.Cheremushkinskaya, 25, Moscow, 117 259, Russia}\\

$\phantom{hj}^{\ddag}$ --
{\it Theory Department,  P. N. Lebedev Physics
Institute, Leninsky prospect, 53, Moscow,\\
{}~117924, Russia}\\

$\phantom{hj}^{\sharp}$ --
{\it Physics Department, Donetsk State University, Donetsk, 340 055,
Ukraine and Donetsk Institute for Physics and Technology, Donetsk,
340 114, Ukraine}}\\\end{quotation}
\end{center}
\bigskip
\bigskip
\centerline{\bf ABSTRACT}
\begin{quotation}
\footnotesize
We demonstrate that the generalization of the relativistic Toda chain (RTC)
is a special reduction of two-dimensional Toda Lattice hierarchy (2DTL).
This reduction implies that the RTC is gauge equivalent to the discrete
AKNS hierarchy
and, which is the same, to the two-component Volterra hierarchy while
its forced (semi-infinite) variant is described by the unitary matrix
integral. The integrable properties of the RTC hierarchy are revealed
in different frameworks of: Lax representation, orthogonal polynomial
systems, and $\tau$-function approach. Relativistic Toda molecule
hierarchy is also considered, along with the forced RTC. Some
applications to biorthogonal polynomial systems are discussed.
\end{quotation}
\end{titlepage}
\clearpage
\newpage

\tableofcontents
\newpage

\section{Introduction}
\setcounter{footnote}{0}
\footnotesize
Since the paper of Ruijsenaars \cite{Rui}, where has been proposed, the
relativistic Toda chain (RTC) system was investigated in many papers
\cite{BR,Sur,KT,Sat}. This system can be defined by
the equation:
\be\label{rt}
 \ddot q_{n} = \;\;(1+{\c}\dot q_{n})(1+{\c}\dot q_{n+1})
\frac{{\rm exp}(q_{n+1}-q_{n})}
{1+{\c^2}{\rm exp}(q_{n+1}-q_{n})} \;\;-\\
 - \;\;(1+{\c}\dot q_{n-1})(1+{\c}\dot q_{n})
\frac{{\rm exp}(q_{n}-q_{n-1})}
{1+{\c^2}{\rm exp}(q_{n}-q_{n-1})}
\ee
which transforms to the ordinary (non-relativistic) Toda chain (TC) in the
evident limit $\c\to 0$. The RTC is integrable, which was discussed
in different frameworks (see, for example, \cite{BR,Sur,KT,Sat} and
references therein). In particular, in \cite{BR} there was introduced a Lax
triad for the RTC, in the paper \cite{Sur} there were investigated Lax
representation and its orbit interpretation for the RTC and the relativistic
Toda molecule, and also suggested a discrete version of the RTC which has the
same integrals of motion (i.e. the same Lax operator, but different evolution).
At last, authors of \cite{Sat} presented the bilinear (Hirota type) equations
whose solutions satisfy the RTC, implying the possibility of $\tau$-function
interpretation of the model.
Thus, the RTC can be obtained from many different starting points.
In particular, this is a limit of the general Ruijsenaars system \cite{Rui}.

On the other hand, it seems that the proper place of the RTC in the framework
of the general theory of the integrable systems has not been adequately
understood yet. Since the seminal papers of Kyoto school \cite{DJKM}
it is evident that any integrable system can be treated  as a particular
reduction of the (multi-component) Kadomtsev-Petviashvili hierarchy. This
statement is certainly true for the RTC too. Using the powerful machinery
of \cite{DJKM}, it is possible to describe wide classes of solutions
to the RTC and reveal many interesting algebraic properties of the system.
Another important lesson of the same philosophy teaches us that many
would-be different systems have the common structures behind if considered
on the level of fundamental notions like Baker-Akhiezer functions,
Grassmannian manifolds etc. In this paper we want to advocate
these points more (in somehow reversed order).
Indeed, we shall show that the RTC has the deep
connection with many well known integrable systems. Namely, the special
case of the AKNS hierarchy \cite{AL} as well as the discrete non-linear
Schr\"odinger equation (of type 2) \cite{AL,FT} are nothing but
different faces of the RTC. Some "novel hierarchy" considered recently
\cite{Rag} is equivalent to RTC also; we shall see that all these systems
belong to the same class of equivalence: the corresponding $L$-operators
are connected with each other by the suitable gauge transformations.
Another interesting observation is that the forced (i.e. semi-infinite)
variant of the RTC is described by the unitary matrix model which, in turn,
naturally arises in the context of the two-dimensional gravity
\cite{PS,GMMMO}.
Actually, the later case was our starting point. Namely, we have started
from the integrable system (with the infinite set of the integrable
flows) \cite{KM} inspired by the unitary matrix model.
The equivalence with the discrete AKNS hierarchy was established there.
Here we show that this system is nothing but the generalized
version of the RTC hierarchy.

The connection with matrix models
\footnote{Let us
stress that the correspondence of the RTC and a unitary matrix model
is not so surprising: it is well-known
that the Hermitean matrix model describes the TC hierarchy \cite{GMMMO}.
On the other hand, the RTC is usually considered as a group generalization
of the TC whatever it means.  Therefore, it is natural to identify the RTC
with some unitary matrix model.}
turns out to be important by some reasons.  First of all, it allows one
to apply effectively the orthogonal polynomials technique, which
considerably simplifies all the derivations and proofs. Indeed, in the
systems of such a type there are two different (but essentially
equivalent) Lax representations -
those given by $2\times2$ matrices depending on the spectral parameter
and given by infinite matrices (in spirit of \cite{UT}). We mostly base on
the second representation, which is effectively treated by orthogonal
polynomials methods (since they solve the corresponding Riemann-Hilbert
problem -- see \cite{GMMMO}).
As a by-product of our consideration,
we obtain some new results specifically interesting for the orthogonal
polynomials theory.
Second, it was shown in \cite{KM} that unitary matrix model is equivalent to
the two-component Volterra system, thus establishing the connection between
this latter and the RTC. On the other hand, Volterra systems are
well investigated
and, therefore, this last connection might help to get more
about the RTC system from this different point of view.

There is another even more useful representation of the RTC.
Namely, we demonstrate in this paper that the RTC is a reduction of the
two-dimensional (subholonomic\footnote{See \cite{UT}.}) Toda lattice
hierarchy (2DTL). On this level, the equivalence of the above mentioned
systems is especially transparent.
Looking at the RTC as a reduction of the 2DTL also allows us to give some
$\tau$-function interpretation of the RTC and realize it in terms of
fermionic correlators. This finally implies the bilinear identities like
those proposed in \cite{Sat} and the theory group
interpretation.  Let us also stress that the $\tau$-function (Grassmannian)
interpretation being absolutely algebraic can be easily deformed to the
quantum case (along the line of the papers \cite{QTF1},\cite{QTF2})
\footnote{Quantization of the RTC was considered in \cite{KT},
where the Sklyanin's scheme of the separation of variables was applied.}.
Furthermore, using group-theoretical approach one can deal with different
RTC's - that is, the general RTC,
forced RTC and RTC molecule hierarchy - from the algebraic point of view.
Their algebraic properties are very interesting. In particular, the
latter hierarchy is related to a class of co-representations of
$U_q(SL(N))$ in the quasi-classical limit. Put it differently,
in the Hamiltonian approach the relativistic Toda molecule
is given as the integrable system on some Poisson leaves on
$SL(N)$ \cite{FM} (see \cite{Soi} for mathematical background).
This is why we consider this approach as a very promising one.

This paper is organized as follows. In section 2 we discuss the simple
spectral problem (three-term recurrent relation) for the RTC which is
natural generalization of the
corresponding problem for the usual Toda Chain (TC). Then, we introduce
two {\it different} flows which lead to the {\it same} equation
(\ref{rt}) and interpret them in the context of the results by
Suris \cite{Sur}. These results, we guess, are not new and can be partly
extracted from preceeding papers (see, for example, \cite{Sur, KT}).
The main aim of section 2 is, however, to establish the key points
which allow us to connect all the integrable structures mentioned above.

Section 3 con\-tains the {\it de\-ri\-vation} of the both RTC
spect\-ral prob\-lem
(proposed in Section 2) and the evolu\-tion equa\-tions from the
(bi)ortho\-gona\-lity con\-ditions implied by the unitary matrix
model. As a by-product of the orthogonality conditions, remarkably,
we automatically get the spectral problem in such a form that
describes the connection of the Baker-Akhiezer functions of the
two-dimensional Toda Lattice hierarchy.
We show that the solution to the RTC hierarchy satisfies simultaneously
the first 2DTL equation with respect two the flows introduced in the
previous section. The connection with the two-component Volterra hierarchy
as well as with the AKNS system is also discussed. As a consequence, we get
the direct correspondence between the RTC, a
particular discrete version of the
non-linear Schr\"odinger equation, and the "novel" integrable system
\cite{Rag}. At the end of the section, we reveal the existence of
{\it two} Lax operators for the RTC which naturally arise from an immediate
generalization of the recurrent relations inspired by the
unitary matrix model.

In section 4 we outline the essential ingredients of the 2DTL theory
\cite{UT} and explicitly describe the whole RTC hierarchy as a particular
reduction of 2DTL.

The forced RTC hierarchy and its finite analog, the relativistic
Toda molecule, are discussed in sections 5 and 6 respectively.
We give the group
interpretation of the later hierarchy. We also discuss the
fermionic representations for the
$\tau$-functions of the corresponding hierarchies and manifestly indicate
the reductions from the 2DTL which describe the hierarchies
in terms of the Grassmannian. We present
also the explicit solution to the forced RTC
in the determinant form.
The fermionic representation received
allows one to make some connection with the results of \cite{Sat}, but
we postpone the discussion of this point till a separate publication
\cite{KMZ} (see, however, section 9).

In section 7 we describe the simple approach to discrete evolutions
of the RTC which is based on the notion of the Darboux-B\"acklund
transformations and can be considered as a natural generalization of the
corresponding notion in the usual Toda chain theory. The continuum limit
as well as the limit to the Toda chain is also discussed here. We consider
also some degeneration of the general Darboux-B\"acklund transformation
leading to simplified discrete equations which can be treated as
a kind of "modified Toda chain" (these equations have been proposed
for the first time in \cite{Sur} but we should stress that they have
nothing to do with the usual discrete-time Toda chain since
the Lax operators of these two systems are different).

Section 8 is devoted to some
applications to the theory of biorthogonal polynomials. Indeed, we discuss
the (relativistic) orthogonal polynomials which satisfy the
three-term recurrent relation specific for the RTC and look at some
very specific solutions to these relations. One of these solutions
describes finite sets of the polynomials corresponding to the Toda
molecule, the other one is
a special limit of the Askey polynomials \cite{as}. In the section, we
discuss the orthogonality measures and some additional
orthogonality relations which can be manifestly treated in the
indicated simple cases.

In concluding remarks we briefly discuss more general (and,
in a sense, more natural) description
of the RTC hierarchy which will
be considered in detail in the forthcoming publication \cite{KMZ}.

In Appendices we give the proof of the some important assertions used
in the main body of the paper. In particular, in Appendix A there were
obtained the evolution equations starting from the orthogonality conditions
determining the unitary matrix model, while in Appendix B the spectral
problem naturally arising in the framework of
orthogonal polynomials is transformed to that of the 2DTL hierarchy.

\sect{Lax representation for RTC}

In this section we describe the Lax representation for the standard RTC
equation. The usual procedure to obtain integrable non-linear equations
consists of the two essential steps:

i)  To find appropriate spectral problem for the Baker-Akhiezer function(s).

ii) To define the proper evolution of this function with respect to
isospectral deformations.

\subsection{Lax representation by three-term recurrent relation}

In the theory of the usual Toda chain the first step implies the
dis\-cre\-tized ver\-sion of the Schr{\" o}\-din\-ger equa\-tion
(see \cite{FT}, for example).
In order to get the relativistic extension of the Toda equations, one should
consider the following "unusual" spectral problem
%\footnote{Generally
%$\Phi_n(z)$ is not a polynomial.}:
\be \label{rec2}
\Phi_{n+1}(z) + a_{n}\Phi_{n}(z) =
z \{\Phi_{n}(z) + b_{n}\Phi_{n-1}(z)\}\;\; ,\;\;\;\;\;n\in\ZZ
\ee
representing a particular discrete Lax
operator acting on the Baker-Akhiezer function $\Phi_n(z)$.  This is a simple
three-term recurrent relation (similar to those for the
Toda and Volterra chains) but with
"unusual" spectral dependence.

As for the second step, one should note that there exist {\it two} distinct
integrable flows leading to the same equation (\ref{rt}). As we shall
see below, the spectral problem (\ref{rec2}) can be naturally incorporated
into the theory of two-dimensional Toda lattice (2DTL) which describes
the evolution with respect to two (infinite) sets of times
$\;\;(t_1, t_2,...)\;\;$,
$\;(t_{-1}, t_{-2},...)\;\;$
(positive and negative times, in accordance with \cite{UT}). In
section 4, we shall derive the general evolution equations satisfied by
$\Phi_{n}$.  Here we describe the two particular flows (at the
moment, we deal with them "by hands", i.e. introducing the
corresponding Lax pairs
by a guess) which lead to the RTC equations (\ref{rt}).
It turns out (as usual) that the simplest equations are to be
associated with the first times of this hierarchy $t_{1}$ and $t_{-1}$.
The most simple evolution equation is
that with respect to the first {\it negative}
time and has the form
\be \label{todaev}
\frac{\d \Phi_{n}}{\d t_{-1}} =
R_{n}\Phi_{n-1}
\ee
with some (yet unknown) $R_{n}$ .  Remarkably enough
that the form of this equation exactly coincides with the corresponding
equation for the non-relativistic Toda chain.

The compatibility condition determines $R_{n}$ in terms of $a_{n}$ and
$b_{n}$:
\be  \label{r}
R_{n} =\frac{b_{n}}{a_{n}}
\ee
and leads to the following equations of motion:
\be \label{a-eq}
\frac{\d a_{n}}{\d t_{-1}} = \frac{b_{n}}{a_{n-1}} -
\frac{b_{n+1}}{a_{n+1}}
\ee
\be  \label{b-eq}
\frac{\d b_{n}}{\d t_{-1}} =
b_{n}\left(\frac{1}{a_{n-1}} - \frac{1}{a_{n}}\right)
\ee
In order to get (\ref{rt}), we should identify
\be\label{a-rep}
a_{n} = \exp(-\c p_{n})
\ee
\be\label{b-rep}
b_{n} = -\c^{2}\exp(q_{n}-q_{n-1}-\c p_{n})
\ee
Note that in this parameterization the "Hamiltonian" $R_{n}$ in (\ref{r})
depends only on coordinates $q_n$'s:
\be\label{RR}
R_{n} = - \c^{2}\exp(q_{n}-q_{n-1})
\ee
Such a form is typical for integrable systems (for example, the first flow
of the TC has the same structure).

In order to preserve the proper limit to the TC equations
as $\c \to 0$, one should perform the rescaling of time in (\ref{a-eq}),
(\ref{b-eq})
\be  \label{resc}
t_{-1} \rightarrow \nu(\c)t_{-1}
\ee
where the function $\nu(\c)$  is to behave as $1/\c$ in the limit
of small $\c$.
Then, from (\ref{a-eq}), (\ref{b-eq})\footnote{Strictly speaking, (\ref{b-eq})
gives the equation  $\frac{\d q_{n}}{\d t_{-1}}-
\frac{\d q_{n-1}}{\d t_{-1}} = Q_{n}-Q_{n-1}$, where $Q_{n}$
is the exponential term in (\ref{q-eq}). Thus, one can only write
$$
\frac{\d q_{n}}{\d t_{-1}} =
-\nu(\c)\{ 1 + \c^{2} \; \exp(q_{n+1}-q_{n})\}
\exp(\c p_{n}) + constant
$$
For the fast-decreasing solutions, $constant=\nu(\c)(1+\c^2)$.
This choice of the constant nicely agrees with the non-relativistic
limit $\c\to 0$:  in this limit $\frac{\d q_{n}}{\d t_{-1}} = p_{n}$.}
\be \label{q-eq}
\frac{\d q_{n}}{\d t_{-1}} =
-\nu(\c)\Bigl\{ 1 + \c^{2} \; \exp(q_{n+1}-q_{n})\Bigr\}
\exp(\c p_{n}) + \nu(\c)(1+\c^{2})
\ee
\be  \label{p-eq}
\frac{\d p_{n}}{\d t_{-1}} =
\c\nu(\c) \Bigl\{ \exp(q_{n} - q_{n-1} + \c p_{n-1}) -
\exp(q_{n+1} - q_{n} + \c p_{n}) \Bigr\}
\ee
After choosing
\be \label{a}
\nu = - [\c(1+\c^{2})]^{-1}
\ee
these equations trivially lead to the RTC equation
\be \label{rtc}
\frac{\d^2 q_{n}}{\d t^2_{-1}} =
\left( 1+\c\frac{\d q_{n+1}}{\d t_{-1}}\right)
\left(1+\c\frac{\d q_{n}}{\d t_{-1}}\right)
\frac{\exp(q_{n+1} - q_{n})}
{1+\c^2\exp(q_{n+1}-q_{n})} \;\;-\\
 - \;\;\left( 1+\c\frac{\d q_{n}}{\d t_{-1}}\right)
 \left( 1+\c\frac{\d q_{n-1}}{\d t_{-1}}\right)
\frac{\exp(q_{n} - q_{n-1})}
{1+\c^2\exp(q_{n}-q_{n-1})}
\ee

Another form of the same equation is given by the replace
\be
\begin{array}{l}
\xi_{n}={\displaystyle \frac{1}{a_{n}}}=\exp(\c p_n)\\
\eta_{n}={\displaystyle \frac{b_{n}}{a_{n}a_{n-1}}}
= - \c^2 \exp(q_n-q_{n-1}+\c p_{n-1})
\end{array}
\ee
Then, from (\ref{a-eq}), (\ref{b-eq})
\footnote{In the paper \cite{KT} $d_{n}=\xi_{n}\;,\;\;
c_{n}=-\eta_{n+1}\;, t=-t_{-1}$ and $\c=1$.}
\be \label{kuz}
\frac{\d \xi_{n}}{\d t_{-1}} = \xi_{n}(\eta_{n+1} -\eta_{n})\\
\frac{\d \eta_{n}}{\d t_{-1}} =
\eta_{n}(\eta_{n+1} - \eta_{n-1} + \xi_{n-1} - \xi_{n})
\ee

As we noted already, the evolution (\ref{todaev}), which leads to the RTC
equations is not the unique one. The other possible choice leading to the same
equations is
\be \label{todaev2}
\frac{\d \Phi_{n}}{\d t_{1}} = - b_{n}(\Phi_{n}-z\Phi_{n-1})
\ee
The compatibility condition of (\ref{rec2}) and (\ref{todaev2}) gives the
equations
\be \label{kuz1}
\frac{\d a_{n}}{\d t_{1}} = - a_{n}(b_{n+1} -b_{n})
\ee
\be \label{kuz2}
\frac{\d b_{n}}{\d t_{1}} = -
b_{n}(b_{n+1} - b_{n-1} + a_{n-1} - a_{n})
\ee
which are exactly of the form (\ref{kuz}). In terms of $(q_{n},\;p_{n})$
we have the following equations (with the same rescaling of time
as in (\ref{resc})):
\be \label{q-eq2}
\frac{\d q_{n}}{\d t_{1}} =
-\nu(\c)\Bigl\{ 1 + \c^{2} \; \exp(q_{n}-q_{n-1})\Bigr\}
\exp(-\c p_{n}) + \nu(\c)(1+\c^{2})
\ee
\be  \label{p-eq2}
\frac{\d p_{n}}{\d t_{1}} = \c\nu(\c) \Bigl\{ \exp(q_{n+1} -
q_{n} - \c p_{n+1}) -
\exp(q_{n} - q_{n-1} - \c p_{n}) \Bigr\}
\ee
slightly different from (\ref{q-eq}), (\ref{p-eq}). Nevertheless,
the second order equation for $q_{n}$ is exactly (\ref{rtc}).
\bigskip

\subsection{$2\times 2$ matrix Lax representation}

The same RTC equation can be obtained from the matrix Lax operator
depending on the spectral parameter \cite{Sur} (generalizing the Lax
operator for the TC \cite{FT}). Then the RTC arises as the compatibility
condition for the following $2\times 2$ matrix equations:
\be
L^{\scriptscriptstyle\rm (S)}_n \psi_n = \psi_{n+1}\;\;\; , \;\;\;
        \frac{\d \psi_{n}}{\d t} = A_{n}\psi_{n}
\ee
where
\be \label{l-op}
L^{\scriptscriptstyle\rm (S)}_{n} = \left(
\begin{array}{cc}
\zeta \exp(\c p_{n}) - \zeta^{-1} &  \c \exp(q_{n})\\
-\c \exp(-q_{n} + \c p_{n})             &  0
\end{array} \right) \;\;\; ; \hspace{5mm}
\psi_{n}=\left(
\begin{array}{c}
\psi^{(1)}_{n}\\
\psi^{(2)}_{n}
\end{array} \right)
\ee
\be \label{m-op}
A_{n} = \left(
\begin{array}{cc}
\c^{2} \exp(q_{n}-q_{n-1} + \c p_{n-1})    &  -\c\zeta^{-1}\exp(q_{n})\\
\c\zeta^{-1}\exp(-q_{n-1} +\c p_{n-1})  & 1 - \; \zeta^{-2}+\c^{2}
\end{array} \right)
\ee
Using (\ref{l-op}), one can re-write the matrix spectral problem as the
recurrent relation for $\psi^{(1)}_{n}$:
\be
\psi^{(1)}_{n+1}(\zeta)=
\bigl\{ \zeta\; \exp(\c p_{n})-\zeta^{-1} \bigr\}
\psi^{(1)}_{n}(\zeta) -
\c^{2}\exp\bigl(q_{n}-q_{n-1}+\c p_{n-1}\bigr)\psi^{(1)}_{n-1}(\zeta)
\ee
Then, putting
\be
\psi^{(1)}_{n}(\zeta) \equiv {\cal N}_{n} \Phi_{n}(\zeta)
\ee
where
\be
\frac{{\cal N}_{n}}{{\cal N}_{n+1}} = \zeta\exp(-\c p_{n})
\ee
one gets
\be  \label{rec1}
\Phi_{n+1}(\zeta) + \exp(-\c p_{n})\Phi_{n}(\zeta) =
\zeta^{2}\bigl\{ \Phi_{n}(\zeta)
- \c^{2} \exp\bigl(q_{n}-q_{n-1}-\c p_{n}\bigr)
\Phi_{n-1}(\zeta)\bigr\}
\ee
This is exactly recurrent relation (\ref{rec2}) with $\;a_n,\;b_n\;$
given by (\ref{a-rep}), (\ref{b-rep}) after the identification
\be \label{spec}
z = \zeta^{2}
\ee
In the same manner, the evolution equation in terms of
$\Phi_{n}$ takes the form
\be
\frac{\d \Phi_{n}}{\d t} = \c^{2}\exp(q_{n}-q_{n-1})\Phi_{n-1}
\ee
Thus, the evolution determined by (\ref{m-op}) is associated with
$\;\;-t_{-1}\;\;$-flow in our approach (see (\ref{todaev}), (\ref{RR})).

Obviously, we are able to construct another $A$-operator which generates
a new integrable flow:
\be \label{m-op2}
A'_{n} = \left(
\begin{array}{cc}
          0   &  -\c\zeta\exp(q_{n}-\c p_{n})\\
\c\zeta\exp(-q_{n-1})  & \hspace{3mm} \zeta^{2}+
                       \c^{2}\exp(q_{n}-q_{n-1}-\c p_{n})-1-\c^{2}
\end{array} \right)
\ee
It is easy to see that (\ref{m-op2}) gives the evolution
equivalent to
(\ref{todaev2}) on the level of the Baker-Akhiezer function $\Phi_n$.

To conclude this section, we should remark that $\;L$-operator (\ref{l-op}),
which determines the RTC is not unique; moreover, it is not the simplest
one. Indeed, in the next section we shall see that there exists the whole
family of the gauge equivalent operators, which contains more "natural"
ones and includes, in particular, the well known operator generating
the AKNS hierarchy. From general point of view, the whole RTC hierarchy is
nothing but AKNS and vice versa.

\sect{RTC and unitary matrix model, AKNS, etc.}

Now we are going to describe the generalized RTC hierarchy as well as its
connection with some other integrable systems. We
start our investigation from the framework of
orthogonal polynomials. The advantage of it is that one does not need to
{\it guess} Lax pair, but instead can {\it get} it automatically using the
formalism developed in \cite{GMMMO,KM,KMMOZ}. From the point of view
of the integrable systems, this method is nothing but the Riemann-Hilbert
problem \cite{GMMMO,KMMOZ}. The orthogonal polynomials give the solutions
only for hierarchies of the forced type (see the proper definition below).
Fortunately, the hierarchies we consider are "local" ones (one-dimensional
subfamilies of the subholonomic systems \cite{UT}) either can be reduced
to these, and, therefore, Lax representations
obtained from the orthogonal polynomials formalism can be continued to the
whole hierarchies \cite{KMMOZ}. From now on, we mostly
arbitrarily turn from the forced hierarchies to the general ones and
back. However, right
in the RTC case, this point still deserves some comments which
we postpone till section 5.

\subsection{Unitary matrix model}

It is well-known that the partition function $\;\tau_n\;$
of the unitary one-matrix model
can be presented as a product of norms of the biorthogonal polynomial
system
\cite{PS,BMS}. Namely, let us introduce a scalar product of the form
\footnote{The signs of positive and negative times are defined in this
way to get the exact correspondence with the times introduced
in \cite{UT}.}
\be\label{sp}
<A,B>=\oint {d\mu(z)\over 2\pi iz}
\exp\Bigl\{\sum_{m>0}(t_mz^m - t_{-m}z^{-m})\Bigr\}A(z)B(\frac{1}{z})
\ee
where the integration measure is not fixed but to be chosen so that
the integral
in (\ref{sp}) is well defined; for example, if $\;d\mu(z)\;$ can be
presented in the form $\;dz\exp(\sum c_kz^k)\;$ then everything reduces
to the usual measure $\;dz\;$ with the proper shifts of times
$t_k\;,\;\;t_{-k}$.
Let us define the system of polynomials biorthogonal with respect to
this scalar product
\be\label{orth-cond}
<\Phi_n,\Phi_k^{\star}>=h_n\delta_{nk}
\ee
Then, the partition function $\;\tau_n\;$of the unitary matrix model is
equal to the product of $h_n$'s:
\be\label{prod}
\tau_n = \prod_{k=0}^{n-1}h_k\;\;,\;\;\;\;\;\;\;\;\tau_0 \equiv 1
\ee
the integration contour in (\ref{sp}) being
the unit circle
\footnote{We do not restrict ourselves to this
integration contour throughout this paper.}
(and, therefore, $h_n$'s are real for the real times $t_k,\;t_{-k}$'s).
The polynomials are
normalized as follows (we should stress that superscript '$\star$' does
not mean the complex conjugation):
\be\label{polyn}
\Phi_n(z)=z^n+\ldots+S_{n-1},\;\;\;\ \
\Phi_n^{\star}(z)=z^n+\ldots+ S^{\star}_{n-1},\;\;\;\; \ \
S_{-1}= S^{\star}_{-1}\equiv 1
\ee

Now it is easy to show that these polynomials satisfy the following
recurrent relations,
%\footnote{This system can be
%easily reduced to two {\it polynomial} equations by introducing the new
%polynomials $\tilde\Phi_n(z)\equiv z^n\Phi^{\star}_n(z^{-1})$.}
which gives the discrete Lax operator:
\be \label{u-rec}
\Phi_{n+1}(z) = z\Phi_{n}(z)+ S_{n}z^{n} \Phi^{\star}_{n}(z^{-1})\\
\Phi^{\star}_{n+1}(z^{-1}) = z^{-1}\Phi^{\star}_{n}(z^{-1})+
S^{\star}_{n}z^{-n}\Phi_{n}(z)
\ee
and
\be\label{hS-rel}
{h_{n+1}\over h_n}=1-S_n S^{\star}_n
\ee

These recurrent relations can be written in several equivalent forms. First, it
can be presented in the form analogous to (\ref{rec2}), i.e. equivalent
to the spectral problem for the RTC:
\be \label{q1}
\Phi_{n+1} - \frac{S_{n}}{S_{n-1}}\Phi_{n}
= z\left\{ \Phi_{n} - \frac{S_{n}}{S_{n-1}}
(1 - S_{n-1}S^{\star}_{n-1})\Phi_{n-1}\right\}
\ee
\be \label{q2}
\Phi^{\star}_{n+1} - \frac{
S^{\star}_{n}}{S^{\star}_{n-1}} \Phi^{\star}_{n} =
z^{-1}\left\{ \Phi^{\star}_{n} - \frac{
S^{\star}_{n}}{S^{\star}_{n-1}}(1 - S_{n-1} S^{\star}_{n-1})
\Phi^{\star}_{n-1}\right\}
\ee
{}From the first relation and using (\ref{rec1}), one can immediately read off
\be\label{Shqp}
{S_n\over S_{n-1}}=-\exp(-\c p_n)\\
{h_n\over h_{n-1}}=-\c^2\exp(q_n-q_{n-1})
\ee
i.e.
\be\label{hq}
h_n=\gamma(-\c^2)^n\exp\bigl( q_n\bigr)
\ee
where $\gamma$ is a constant which does not depend on $n$.
%This constant
%practically does not effect to the system and can be removed, for
%example, by imposing on $h_n$ the fast-decreasing boundary conditions.
Thus, the orthogonality conditions (\ref{orth-cond}) lead exactly to
the spectral problem for the RTC. We should stress that equations
(\ref{q1}), (\ref{q2}) can be {\it derived} from system (\ref{u-rec}).
Usually, the three-term recurrent relations can be transformed to the
system of two equations in many different ways (see below).
The main feature of (\ref{u-rec}), which distinguishes it from the
other possible choices
is the remarkable fact: this system describes the connection
between the pair of the Baker-Akhiezer functions arising in the context
of 2DTL. We shall discuss this point in section 4.

Some remarks are in order now. In all formulas above the discrete index
$\;\;n\;\;$ runs over the non-negative integers. One can trivially extend
all relations (\ref{u-rec})-(\ref{hq}) (and all the relations below)
to arbitrary $\;n\in\ZZ\;$. Then, the polynomial case corresponds to
the conditions
\be
\frac{h_n}{h_{n-1}} \equiv 0\;\;, \;\;\;\; n<0
\ee
or, (see (\ref{Shqp}))
\be
q_{-1} = \infty
\ee
with (formal) ordering $\;q_{-1} < q_{-2} < \;\ldots\;$ .
Equivalently, the analytical continuation of (\ref{prod}) written
in the form
\be \label{tau}
h_n = \frac{\tau_{n+1}}{\tau_n}
\ee
reduces to the polynomial case by imposing the conditions
\be
\tau_n \equiv 0\;\;, \;\;\;\; n\;<0
\ee
This is exactly what we call the {\it forced} integrable
systems \cite{KMMOZ}.
Our way of doing is to derive all the formulas in the
forced ($\equiv$ polynomial) case and then to extend them to
arbitrary values of the discrete indices.

Using the orthogonal conditions, it is also possible to obtain the equations
which describe the time dependence of $\;\Phi_n,\;\Phi^{\star}_n\;$.
Namely, differentiating (\ref{orth-cond}) with respect to times
$\;t_1\;,\;\;t_{-1}\;$ gives the following evolution equations
(for derivation see Appendix A):
\be\label{qt1}
\frac{\d \Phi_{n}}{\d t_{1}}=
\frac{S_{n}}{S_{n-1}}\frac{h_{n}}{h_{n-1}}
(\Phi_{n} -z \Phi_{n-1})
\ee
\be \label{qt2}
\frac{\d \Phi_{n}}{\d t_{-1}} =
\frac{h_{n}}{h_{n-1}}\Phi_{n-1}
\ee
\be \label{barqt1}
\frac{\d \Phi^{\star}_{n}}{\d t_{1}} =  -
\frac{h_{n}}{h_{n-1}}\Phi^{\star}_{n-1}
\ee
\be\label{barqt2}
\frac{\d \Phi^{\star}_{n}}{\d t_{-1}}= -
\frac{S^{\star}_{n}}{S^{\star}_{n-1}}
\frac{h_{n}}{h_{n-1}}(\Phi^{\star}_{n} -z^{-1}\Phi^{\star}_{n-1})
\ee
(see general evolution equations with respect to higher flows in the next
section).
The compatibility conditions give the following nonlinear
evolution equations
\footnote{ In the polynomial case, one can get these equations directly
from (\ref{qt1})-(\ref{barqt2}) considering terms of the order $\;\sim z^0$;
see definition (\ref{polyn}). }:
\be  \label{st1} \frac{\d S_{n}}{\d t_{1}} =
S_{n+1}\frac{h_{n+1}}{h_{n}}
\ee
\be  \label{st2}
\frac{\d S_{n}}{\d t_{-1}} = S_{n-1}\frac{h_{n+1}}{h_{n}}
\ee
\be  \label{barst1}
\frac{\d S^{\star}_{n}}{\d t_{1}} = - S^{\star}_{n-1}\frac{h_{n+1}}{h_{n}}
\ee
\be  \label{barst2}
\frac{\d S^{\star}_{n}}{\d t_{-1}} = - S^{\star}_{n+1}\frac{h_{n+1}}{h_{n}}
\ee
As a consequence, in the polynomial case,
\be  \label{ht1}
\frac{\d h_{n}}{\d t_{1}} = - S_{n}S^{\star}_{n-1}h_n
\ee
\be \label{ht2}
\frac{\d h_{n}}{\d t_{-1}} = S_{n-1}S^{\star}_{n}h_n
\ee
These are exactly relativistic Toda equations written in somewhat
different form.
Indeed, from (\ref{ht1}), (\ref{st1}) and (\ref{barst1}) one gets
\be\label{h-eq}
\frac{\d^2}{\d t_1^2}\log h_n = - S_{n+1}S^{\star}_{n-1}\frac{h_{n+1}}{h_n}+
S_{n}S^{\star}_{n-2}\frac{h_{n}}{h_{n-1}}
\ee
Using (\ref{ht1}) again and (\ref{hS-rel})
\be
\left(\frac{\d}{\d t_1}\log h_{n}\right)
\left(\frac{\d}{\d t_1} \log h_{n-1}\right) =
S_{n}S^{\star}_{n-2}\left(1-\frac{h_n}{h_{n-1}}\right)
\ee
Substitution of $\;S_{n}S^{\star}_{n-2}\;$ to (\ref{h-eq}) gives
\footnote{The same equation holds for $t_{-1}$-flow.}
\be\label{h-eqs}
\frac{\d^2}{\d t_1^2}\log h_n = - \left(\frac{\d}{\d t_1}\log h_{n}\right)
\left(\frac{\d}{\d t_1}\log h_{n+1}\right)
\frac{{\displaystyle \frac{h_{n+1}}{h_n}}}
{\displaystyle {1-\frac{h_{n+1}}{h_n}}}\; + \\
+\; \left(\frac{\d}{\d t_1}\log h_{n-1}\right)
\left(\frac{\d}{\d t_1}\log h_{n}\right)
\frac{{\displaystyle \frac{h_{n}}{h_{n-1}}}}
{\displaystyle {1-\frac{h_{n}}{h_{n-1}}}}
\ee
On the other hand, the RTC is a particular case of the 2DTL hierarchy.
Indeed, let us introduce the key objects in the theory of integrable
systems - the $\tau$-function as it is defined in (\ref{tau}).
Then, with the help of (\ref{st1})-(\ref{ht2}), one can show that
the functions
$\;\tau_n\;$ satisfy the first equation of the 2DTL:
\be\label{todaeq}
\d_{t_1}\d_{t_{-1}}\log \tau_n = -\frac{\tau_{n+1}\tau_{n-1}}
{\tau^2_{n}}
\ee
Therefore, it is natural to assume that the higher flows generate the
whole set of non-linear equations of the 2DTL in spirit of \cite{UT}.
In the next section, we shall see that this is indeed the case. On the
other hand, it turns out that the integrable system determined by
(\ref{q1}), (\ref{q2}) is highly degenerate as compared with the general
2DTL. Indeed, the solutions to the equations (\ref{q1}), (\ref{q2})
(treated for a moment as two separate equations) correspond to the four
Baker-Akhiezer functions of the 2DTL. In the general theory \cite{UT}, these
functions are linearly independent. In the RTC case, they are
dependent due to (\ref{u-rec}); this is the origin of the degeneracy.
Actually, this degeneracy is responsible for the appearance of additional
non-linear equations, which are absent in typical situation. For example,
these are equations (\ref{h-eqs}), which contain the
derivatives of $\;\log h_n\;$ with respect to $\;t_1\;$ (or $\;t_{-1}$)
only. These additional equations being consistent with the whole 2DTL
hierarchy specialize the reduction of the 2DTL.
Therefore, we should treat the RTC hierarchy as
a special reduction of an "abstract" 2DTL (see section 4).
\footnote{Let us stress that
the term "reduction" does not obligatory mean the polynomial case since
non-polynomial solutions to (\ref{q1}), (\ref{q2}), (\ref{qt1})-
(\ref{barqt2}) are also reduced in the above mentioned sense.}

Let us also note that, using (\ref{hq}), one can get from
equation (\ref{h-eqs}) the equation distinguished from (\ref{rt}) by
the linear (in time) shift of the variable $q_n\to q_n-{1\over \epsilon}t_1$.
This means that one should careful when express $q_n$ through
$\tau$-functions.

This completes the derivation of the RTC from the unitary matrix model.
Now we consider the connection of the RTC with the integrable systems
mentioned in the Introduction.

\subsection{Two-component Volterra}

Recursion relations (\ref{q1})--(\ref{q2}) can be presented in the form
establishing their connection with the two-component Volterra hierarchy
\cite{KM} (it should be understood that, generally, $\Phi_n$ and
$\Phi_n^{\star}$ in these relations are not polynomials).
For doing this, we introduce the two-component {\it functions}
\be\label{LaxV}
f^{(\pm)}_n(z)  \equiv  z^{-{n\over 2}+1}\Phi_{n-1}(z)\pm
z^{{n\over 2}-1} \Phi^{\star}_{n-1}(z^{-1})
\ee
These functions satisfy the following orthogonality conditions:
\be
<f^{(\pm)}_n,f^{(\pm)}_k>=\left[2\mp(S_{n-1}+
S^{\star}_{n-1})\right]h_{n-1}\delta_{nk}\\
<f^{(+)}_n,f^{(-)}_k>=(S_{n-1}-S^{\star}_{n-1})h_{n-1}\delta_{nk}
\ee
These functions also satisfy the recurrent
relations which can be written in the compact matrix form after introducing
the two-component column
\be
f_n\equiv \left(
\begin{array}{c}
f^{(+)}_n\\f^{(-)}_n
\end{array}
\right)
\ee
Then, one gets the recurrent relations giving the two-component
Volterra Lax operator
\be\label{LaxVolt}
(z^{{\f 2}}+z^{-{\f 2}})f_n = f_{n+1} + V_nf_{n-1}\\
(z^{{\f 2}}-z^{-{\f 2}})f_n = \sigma_1f_{n+1} +
\tilde V_nf_{n-1}
\ee
where
\be V_n\equiv W_nW_{n-1}^{-1}\;\; ,\ \ \
\tilde V_n\equiv W_n\sigma_3 W_{n-1}^{-1}
\ee
$\sigma_i$'s are Pauli matrices and
\be
W_n=h_{n-1}\left(
\begin{array}{cc}
1-S_{n-1} & 1- S^{\star}_{n-1}\\
-(1+S_{n-1}) & 1+ S^{\star}_{n-1}
\end{array}
\right)\ ;\ \ \det W_n=2h_n h_{n-1}
\ee
Three forms of the recurrent relations (\ref{u-rec}), (\ref{q1}) and
(\ref{LaxVolt})
explicitly establish the connection between the unitary
matrix model, RTC and two-component Volterra system
(certainly, there still remains to compare the evolutions, see
the next section).

\subsection{RTC versus AKNS and "novel" hierarchies}

Now let us demonstrate the correspondence between RTC and AKNS system.
We have already seen that the orthogonality conditions naturally lead
to the 2$\times$2 formulation of the problem generated by the unitary matrix
model:
\be\label{U1}
L^{\scriptscriptstyle\rm(U)}\left(
\begin{array}{c}
\Phi_n\\
\Phi^{\star}_n
\end{array}\right) = \left(
\begin{array}{c}
\Phi_{n+1}\\
\Phi^{\star}_{n+1}
\end{array}\right)\;\; ,
\;\;\;\;\;
L^{\scriptscriptstyle\rm(U)}=\left(
\begin{array}{cc}
    z             &  \;\;\;\; z^nS_n\\
z^{-n}S_n^{\star} &  \;\;\;\; z^{-1}
\end{array}\right)
\ee
\be\label{U2}
\frac{\d}{d t_1}\left(
\begin{array}{c}
\Phi_n\\
\Phi^{\star}_n
\end{array}\right)=\left(
\begin{array}{cc}
-S_nS_{n-1}^{\star} & \;\;\;z^nS_n\\
z^{1-n}S^{\star}_{n-1}  & \;\;\;-z
\end{array}\right)\left(
\begin{array}{c}
\Phi_n\\
\Phi^{\star}_n
\end{array}\right)
\ee
\be\label{U3}
\frac{\d}{d t_{-1}}\left(
\begin{array}{c}
\Phi_n\\
\Phi^{\star}_n
\end{array}\right)=\left(
\begin{array}{cc}
       z^{-1}             & \;\;\; -z^{n-1}S_{n-1}\\
-z^{-n}S^{\star}_{n}  & \;\;\; S_{n-1}S_{n}^{\star}
\end{array}\right)\left(
\begin{array}{c}
\Phi_n\\
\Phi^{\star}_n
\end{array}\right)
\ee
(Equations (\ref{U2}), (\ref{U3}) follow from (\ref{qt1})-(\ref{barqt2})
and the original spectral problem (\ref{u-rec})).
Put
\be  \label{qp}
\Phi_{n} \equiv z^{n/2-1/4}F_{n}  \\
\Phi^{\star}_{n} \equiv z^{-n/2+1/4}F^{\star}_{n}
\ee
Then the spectral problem (\ref{U1}) can be rewritten in the matrix form
\be
L_{n}^{\scriptscriptstyle\rm (AKNS)}
{\cal F}_{n} = {\cal F}_{n+1} \;\; , \;\;\;\;\;
{\cal F}_{n} \equiv \left(
\begin{array}{l}
F_{n}\\
F^{\star}_{n}
\end{array} \right)
\ee
where
\be \label{akns}
L_{n}^{\scriptscriptstyle\rm (AKNS)} = \left(
\begin{array}{ll}
\zeta           &   S_{n}\\
S^{\star}_{n}   &   \zeta^{-1}
\end{array} \right) \;\;, \;\;\;\;\;\; \zeta \equiv z^{1/2}
\ee
This is the Lax operator for the discrete AKNS \cite{AL}.
Obviously the evolution equations (\ref{U2}), (\ref{U3}) can be written
in terms of $F_{n},\;
F^{\star}_{n}$ as
\be
\frac{\d {\cal F}_{n}}{\d t_{1}} = A^{(1)}_{n}{\cal F}_{n}\;\; ,
\hspace{12mm} A^{(1)}_{n} = \left(
\begin{array}{ll}
- S_{n}S^{\star}_{n-1}        &   \zeta S_{n}\\
\zeta S^{\star}_{n-1}       &   -\zeta^{2}
\end{array} \right)
\ee
\be
\frac{\d {\cal F}_{n}}{\d t_{-1}} =\;-\;
A^{(-1)}_{n}{\cal F}_{n}\;\; ,
\hspace{12mm} A^{(-1)}_{n} = \left(
\begin{array}{ll}
-\zeta^{-2}              &   \zeta^{-1} S_{n-1}\\
\zeta^{-1} S^{\star}_{n}    &   - S_{n-1}S^{\star}_{n}
\end{array} \right)
\ee
Note that after introducing the trivial flow
\be
\frac{\d {\cal F}_{n}}{\d t_{0}} =
A^{(0)}_{n}{\cal F}_{n}\;\; ,
 \hspace{12mm} A^{(0)}_{n} = \left(
\begin{array}{ll}
1  &  0 \\
0  &  -1
\end{array} \right)
\ee
we get the difference non-linear Schr{\" o}dinger system (DNLS)
\cite{AL} (see also \cite{FT}) generated by the "mixed" flow
\be\label{T-dep}
\frac{\d {\cal F}_{n}}{\d T} \equiv
\left( \frac{\d}{\d t_{0}}  - \frac{\d}{\d t_{-1}}
- \frac{\d}{\d t_{1}}\right) {\cal F}_{n}
= (A^{(0)}_{n} + A^{(-1)}_{n} - A^{(1)}_{n}){\cal F}_{n}\equiv\\
\equiv\;\;\left(
\begin{array}{ll}
1+S_nS_{n-1}^{\star} -\zeta^{-2}\;\;\;\; & \zeta^{-1}S_{n-1} - \zeta S_n \\
\zeta^{-1}S_{n}^{\star} - \zeta S_{n-1}^{\star} \;\;\;\;&
-1 - S_{n-1}S_n^{\star} +\zeta^2
\end{array}\right)
\ee
Indeed, from the compatibility conditions for (\ref{akns}), (\ref{T-dep})
or, equivalently, just from (\ref{st1})-(\ref{st2})  (along with the trivial
evolution $
\d_{t_0} S_n = 2S_n\;,\;
\d_{t_0} S^{\star}_n= -2S^{\star}_n\;$)
one gets the discrete version of the nonlinear Schr\"odinger equation:
\be
\frac{\d S_n}{\d T} = - (S_{n+1} - 2S_n + S_{n-1}) + S_nS_n^{\star}
\bigl( S_{n+1} + S_{n-1}\bigr)
\ee
Note also that the "novel" hierarchy of \cite{Rag} is equivalent to the RTC
(and, therefore, to the AKNS hierarchy) as well. Namely, the Lax operator in
\cite{Rag}, i.e.
\be \label{new}
{\widehat L}_{n} = \left(
\begin{array}{cc}
z + u_{n}v_{n}    &  \;\;\;  u_{n}\\
     v_{n}        &   1
\end{array} \right) \;\; ; \;\;\;\;\;\;\;\;\;\;\;
{\widehat L}_n\left(
\begin{array}{l}
\phi_n^{(1)}\\
\phi_n^{(2)}
\end{array}\right) =   \left(
\begin{array}{l}
\phi_{n+1}^{(1)}\\
\phi_{n+1}^{(2)}
\end{array}\right)
\ee
defines the recurrent relation of the form (\ref{q1}):
\be \label{rag1}
\phi^{(1)}_{n+1} - \left( u_{n}v_{n} + \frac{u_{n}}{u_{n-1}}\right)
 \phi^{(1)}_{n} = z \left( \phi^{(1)}_{n} -
 \frac{u_{n}}{u_{n-1}}\phi^{(1)}_{n-1} \right)
\ee
thus revealing the connection with the RTC.
Comparing (\ref{q1}) and (\ref{rag1}) leads to the
identification
\be\label{u-v}
u_{n} = S_{n}h_{n} \;\;\; ,\hspace{10mm}
v_{n} = \frac{S^{\star}_{n-1}}{h_{n}}
\ee
where $h_{n}$'s satisfy (\ref{hS-rel}). Moreover, from (\ref{new})
and (\ref{u-rec}) it is easy to see that
\be
\phi^{(1)}_n = \Phi_n\\
\phi^{(2)}_n = \frac{1}{h_n}\Bigl(z^n\Phi^{\star}_n -
S^{\star}_{n-1}\Phi_n\Bigr)
\ee
and, therefore, $\;{\widehat L}_n\;$ can be obtained from
$\;L_{n}^{\scriptscriptstyle\rm (AKNS)}\;$ by the discrete gauge
transformation:
\be
{\widehat L}_{n} =
U_{n+1}L_{n}^{\scriptscriptstyle\rm (AKNS)}U_{n}^{-1}
\ee
(for $z = \zeta^{2}$) where
\be
U_n = z^{n/2-1/4}\left(
\begin{array}{cc}
1\; & 0\\
-\frac{S_{n-1}^{\star}}{h_n} & \frac{z^{1/2}}{h_n}
\end{array}\right)
\ee
Evolution equations (\ref{st1})-
(\ref{ht2}) in terms of new variables (\ref{u-v}) have the form
\be
\begin{array}{ll}
\frac{\d u_n}{\d t_1} = u_{n+1} - u_n^2v_n \;\; , \;\;\;\;\;\;\;\; &
\frac{\d u_n}{\d t_{-1}} = \frac{u_{n-1}}{1+ u_{n-1}v_n}\\
\frac{\d v_n}{\d t_1} = - v_{n-1} + u_nv_n^2 \;\; , \;\;\;\;\;\;\;\; &
\frac{\d v_n}{\d t_{-1}} = - \frac{v_{n+1}}{1+ u_{n}v_{n+1}}
\end{array}
\ee
and easily reproduce the usual AKNS equations in the continuum
limit since
\be
\bigl(\d_{t_{0}} - \d_{t_{1}}-\d_{t_{-1}}\bigr)u_n  =
-\bigl(u_{n+1} -2u_n + u_{n-1}\bigr) +
\bigl(u^2_{n-1}+u^2_n\bigr)v_n + \ldots \\
\bigl(\d_{t_{0}} - \d_{t_{1}}-\d_{t_{-1}}\bigr)v_n  =
\bigl(v_{n+1} -2v_n + v_{n-1}\bigr) -
\bigl(v^2_{n}+v^2_{n+1}\bigr)u_n + \ldots
\ee
We conclude with
the remark that the operator $\;L_n^{\scriptscriptstyle\rm (S)}\;$
in (\ref{l-op}) is also gauge equivalent to
$\;L_{n}^{\scriptscriptstyle\rm (AKNS)}\;$:
\be
L_{n}^{\scriptscriptstyle\rm (S)} =
{\widetilde U}_{n+1}L_{n}^{\scriptscriptstyle\rm (AKNS)}
{\widetilde U}_{n}^{-1}
\ee
where
\be
{\widetilde U}_n = \left(
\begin{array}{cc}
\frac{(-1)^{n}}{S_{n-1}}\; & 0\\
\frac{\c^{2n-1} z^{-1/2}}{S_{n-1}h_n}  &
- \frac{\c^{2n-1}}{h_n}
\end{array}\right)
\ee

\subsection{Non-local Lax representation}

At the end of the section let us note that there is another form of the
recurrent relations which is non-local (i.e. contains all the functions with
smaller indices) but instead expresses  $\Phi_n(z)$ through
themselves. This form of the spectral problem will be crucial for dealing
with the RTC as a particular reduction of the 2DTL.
Let us introduce the normalized functions
\be\label{PP}
{\cal P}_n(z) \equiv  \Phi_n(z)\ \;\;,\;\;\;\;\;\;
        {\cal P}^{\star}_n(z^{-1}) \equiv  {1\over h_n}
    \Phi^{\star}_n(z^{-1})
\ee
such that, for example, in the polynomial case
\be
\langle {\cal P}_n, {\cal P}^{\star}_k\rangle  =
\delta _{nk}
\ee
Then, from (\ref{q1}), (\ref{q2}) one can show
that in the forced and fast-decreasing cases some proper solutions
(see the discussion in Appendix B) satisfy the equations
\be\label{Lax2DTL}
z{\cal P}_n(z) =
{\cal P}_{n+1}(z) - S_nh_n\sum ^n_{k=-\infty} {S^{\star}_{k-1}\over h_k}
{\cal P}_k(z) \equiv  {\cal L}_{nk}{\cal P}_k(z)\\
z^{-1}{\cal P}^{\star}_n(z^{-1}) =
{h_{n+1}\over h_n} {\cal P}^{\star}_{n+1}(z^{-1}) - S^{\star}_n
\sum^n_{k=-\infty}S_{k-1} {\cal P}^{\star}_k(z^{-1}) \equiv
\ov{\cal L}_{kn} {\cal P}^{\star}_k(z^{-1})
\ee
Note that this expression is correct for
general (non-polynomial) ${\cal P}_n$ and ${\cal P}^{\star}_n$ provided the
sums run over all integer $k$. In the polynomial case, the sums automatically
run over only non-negative $k$. Let us also stress
the natural appearance of
variables (\ref{u-v}) in the Lax operator $\;{\cal L}_{nk}\;$.
The last representation of the spectral problem will be useful in the
next section to determine the general evolution of the system. Indeed,
these relations manifestly describe the embedding of the RTC into the
2DTL \cite{UT,KM}, which is given essentially by {\it two} Lax operators
(${\cal L}$ and $\ov{\cal L}$).

\sect{RTC as reduction of 2DTL}

\subsection{Hierarchy evolution}

In order to determine the whole set of the evolution equations, one can
use different tricks. The simplest one is to use the orthogonal polynomial
technique and then to continue the result for the "general" hierarchy.
We use this way at the end of the section. Now let us note that
there is also some less direct way to obtain the evolution
suitable even in the non-polynomial case without continuation.
Namely, one can use embedding (\ref{Lax2DTL}) of the system
into the 2DTL, making
use of the standard evolution of this latter \cite{UT}. Let us briefly
describe the formalism of the 2DTL following \cite{UT}. In their framework,
one introduces {\it two} different Baker-Akhiezer (BA) $\ZZ\times\ZZ$
matrices ${\cal W}$ and $\ov{\cal W}$. (In the original paper \cite{UT},
these matrices were
denoted as $W^{(\infty)}$ and $W^{(0)}$ ;
the superscripts correspond to the marked
points on the Riemann surface which specify the solution to the 2DTL,
i.e. the point of the infinite dimensional Grassmannian).
These matrices satisfy the linear system:\\
\phantom{1cm}i) the matrix version of the spectral problem:
\be\label{OE2}
{\cal L}{\cal W} = {\cal W} \Lambda\;\;,\ \ \
\ov{{\cal L}}\;\ov{\cal W} = \ov{\cal W}\Lambda ^{-1}
\ee
\phantom{1cm}ii) the matrix version of the evolution equations:
\be\label{OE3}
\begin{array}{lll}
\frac{\d {\cal W}}{\d t_{m}}=({\cal L}^{m})_{+}{\cal W}\;,\;\;\;&
\frac{\d \ov{{\cal W}}}{\d t_{m}}=
({\cal L}^{m})_{+}\ov{{\cal W}}&\\
\frac{\d {\cal W}}{\d t_{-m}}=(\ov{\cal L}^{\;m})_{-}{\cal W}\;,\;\;\;&
\frac{\d \ov{{\cal W}}}{\d t_{-m}}=(\ov{\cal L}^{\;m})_{-}\ov{{\cal W}}
\;,\;\;\;&m=1,2,...
\end{array}
\ee
where $\ZZ\times\ZZ$ matrices ${\cal L}$ and $\ov{\cal L}$ have
(by definition) the following structure:
\be \label{lax1}
{\cal L} = \sum_{i\leq 1}{\rm diag}[b_{i}(s)]\Lambda^i \;\; ;\;\;\;
	b_{1}(s)=1\\
\ov{\cal L} = \sum_{i\geq -1}{\rm diag}[c_{i}(s)]\Lambda^i \;\; ;\;\;\;
	c_{-1}(s)\neq 0
\ee
Here ${\rm diag}[b_{i}(s)]$ denotes an infinite diagonal matrix
$\;{\rm diag}(\ldots \;b_{i}(-1),\; b_{i}(0), \;b_{i}(1),\;\ldots$) ;$\;\;\;$
$\Lambda$ is the shift matrix with the elements $\Lambda_{nk}
\equiv \delta_{n,k-1}$ and for arbitrary infinite matrix
$A=\sum_{i\in\ZZ}{\rm diag}[a_{i}(s)]\Lambda^{i}$ we set
\be
(A)_{+}\equiv\sum_{i\geq 0}{\rm diag}[a_{i}(s)]\Lambda^{i}\;\;,\;\;\;\;\;
(A)_{-}\equiv\sum_{i<0}{\rm diag}[a_{i}(s)]\Lambda^{i}
\ee
i.e. $(A)_{+}$ is the upper triangular part of the matrix $A$ (including
the main diagonal) while $(A)_{-}$ is strictly the lower triangular part.

Note that (\ref{lax1}) can be written in components as
\be\label{lax2}
{\cal L}_{nk}=\delta_{n+1,k} +
b_{k-n}(n)\theta (n-k)\;,\;\;\;
\ov{\cal L}_{nk}=c_{-1}(n)\delta_{n-1,k} +
c_{k-n}(n)\theta (k-n) \;\; ; \;\;\;\;n,k\in\ZZ
\ee
The compatibility conditions
imposed on (\ref{OE2}),(\ref{OE3}) give rise
to the infinite set (hierarchy) of
nonlinear equations for the operators ${\cal L}$,
$\ov{\cal L}$ or, equivalently, for the coefficients $b_{m}(n)$, $c_{m}(n)$.
This is what is called 2DTL hierarchy.

On the level of nonlinear equations, one does not need the information
of the structure of BA matrices. However, in order to get the
touch with the polynomials, this information is essential.
It was proved in \cite{UT} that the linear system (\ref{OE2}),
(\ref{OE3}) is resolved by the following BA matrices:
\be
{\cal W} =  {\rm\bf W}
\exp \left\lbrace \sum _{m=1}^{\infty}t_m\Lambda^m\right\rbrace\;\;, \;\;\;
\ov{{\cal W}} = \ov{{\rm\bf W}}
\exp \left\lbrace \sum _{m=1}^{\infty}t_{-m}\Lambda ^{-m}\right\rbrace
\ee
where ${\rm\bf W}$, $\ov{\rm\bf W}$ can be presented in the form:
\be
{\rm\bf W} \equiv  \sum _{i=0}^{\infty}\hbox{
diag}[w_{i}(s)] \Lambda^{-i}\;,\;\;
{\rm\bf W}^{-1} \equiv  \sum _{i=0}^{\infty}\Lambda^{-i}\hbox{
diag}[w_{i}^{\star}(s+1)] \;,\;\;\;w_{0}(s) = w_{0}^{\star}(s) \equiv 1
\ee
\be
\ov{\rm\bf W} \equiv  \sum _{i=0}^{\infty}\hbox{
diag}[\ov{w}_{i}(s)] \Lambda^{i}\;,\;\;
\ov{\rm\bf W}\;^{-1} \equiv  \sum _{i=0}^{\infty} \Lambda^{i}
\hbox{diag}[\ov{w}_{i}^{\;\star}(s+1)]
\ee
Now let us introduce the BA {\it functions} as follows:
\be\label{w1}
w_{n}(z) \equiv \sum_{k\in\ZZ}{\rm\bf W}_{nk}z^{k}=
z^{n}\sum_{i=0}^{\infty}w_{i}(n)z^{-i}
\ee
\be\label{w2}
\ov{w}_{n}(z) \equiv \sum_{k\in\ZZ}\ov{\rm\bf W}_{nk}z^{k}=
z^{n}\sum_{i=0}^{\infty}\ov{w}_{i}(n)z^{i}
\ee
At the same time, we define the {\it adjoint} BA functions as
\be\label{ad1}
zw_{n+1}^{\star}(z) \equiv \sum_{k\in\ZZ}{\rm\bf W}^{-1}_{kn}z^{-k}
\;\; ;\;\;\;\;
w_{n}^{\star}(z)=z^{-n}\sum_{i=0}^{\infty}w_{i}^{\star}(n)z^{-i}
\ee
\be\label{ad2}
z\ov{w}_{n+1}^{\;\star}(z) \equiv \sum_{k\in\ZZ}
\ov{\rm\bf W}^{\;-1}_{kn}z^{-k}\;\; ;
\;\;\;\;\; \ov{w}_{n}^{\;\star}(z)=
z^{-n}\sum_{i=0}^{\infty}\ov{w}_{i}^{\;\star}(n)z^{i}
\ee
Using (\ref{OE2}), (\ref{OE3}), it is easy to show
that the BA functions satisfy the linear equations:
\be\label{laxTL}
{\cal L}_{nk}w_{k}(z) = zw_{n}(z)\;\;,\;\;\;
\ov{\cal L}_{nk}\ov{w}_{k}(z) = z^{-1}\ov{w}_{n}(z)
\ee \be\label{OE4}
\begin{array}{lll}
\frac{\d
w_{n}(z)}{\d t_{m}}=-[({\cal L}^{m})_{-}]_{nk}w_{k}(z)\;,\;\;\;&
\frac{\d\ov{w}_{n}(z)}{\d t_{m}}=
[({\cal L}^{m})_{+}]_{nk} \ov{w}_{k}(z)&\\
\frac{\d w_{n}(z)}{\d t_{-m}}=
[(\ov{\cal L}^{\;m})_{-}]_{nk}w_{k}(z)\;,\;\;\;&
\frac{\d \ov{w}_{n}}{\d t_{-m}}=
-[(\ov{\cal L}^{\;m})_{+}]_{nk}\ov{w}_{k}(z)
\;,\;\;\;&m=1,2,...
\end{array}
\ee
Corresponding system for the adjoint BA functions is:
\be\label{adLax}
{\cal L}_{kn}w_{k+1}^{\star}(z) = zw_{n+1}^{\star}(z)\;\;,\;\;\;
\ov{\cal L}_{kn}\ov{w}_{k+1}^{\;\star}(z) =
z^{-1}\ov{w}_{n+1}^{\;\star}(z)
\ee
\be\label{OE5}
\begin{array}{lll}
\frac{\d w_{n+1}^{\star}(z)}{\d t_{m}}=[({\cal L}^{m})_{-}]_{kn}
w_{k+1}^{\star}(z)\;,\;\;\;&
\frac{\d \ov{w}_{n+1}^{\;\star}(z)}{\d t_{m}}=-[({\cal L}^{m})_{+}]_{kn}
\ov{w}_{k+1}^{\;\star}(z) &\\
\frac{\d w_{n+1}^{\star}(z)}{\d t_{-m}}=-[(\ov{\cal L}^{\;m})_{-}]_{nk}
w_{k+1}^{\star}(z)\;,\;\;\;&
\frac{\d \ov{w}_{n+1}^{\;\star}}{\d t_{-m}}=[(\ov{\cal L}^{\;m})_{+}]_{kn}
\ov{w}_{k+1}^{\;\star}(z)
\;,\;\;\;&m=1,2,...
\end{array}
\ee

\subsection{Reduction from 2DTL}

Now let us return to the case of RTC "in general situation", i.e. with the
conditions $S_n =1 , S^{\star}_n =1 \;\;n<0$ restricting onto
the polynomial case being removed away.
{}From (\ref{Lax2DTL}), one gets two matrices
\be\label{RTC1}
{\cal L}_{nk} = \delta_{n+1,k} -\frac{h_n}{h_k}S_{n}S^{\star}_{k-1}
\theta(n-k)\;\;,\;\;\;\;\;k, n \in \ZZ
\ee
\be\label{RTC2}
\ov{\cal L}_{nk} = \frac{h_n}{h_{n-1}}\delta_{n-1,k}
-S_{n-1}S^{\star}_{k}\theta(k-n)\;\;,\;\;\;\;\;k, n \in \ZZ
\ee
which have exactly the form (\ref{lax2}). One can consider the spectral
problem (\ref{laxTL}) for these particular operators. It is easy to prove
that every solution to (\ref{laxTL}), (\ref{RTC1}), (\ref{RTC2}) satisfies
the recurrent equation (\ref{q1}). On the other hand, (\ref{q1}) has two
linear independent solutions with asymptotics (we are using the
identification (\ref{PP}))
\be \label{p1}
{\cal P}^{(1)}_{n}(z) = z^{n}\Bigl(1 + O\Bigl(\frac{1}{z}\Bigr)\Bigr)
\;\;\;\;\; z\to\infty
\ee
\be \label{p2}
{\cal P}^{(2)}_{n}(z) = h_n z^{n}\Bigl(1 + O(z)\Bigr)\;\;\;\;\;z\to 0
\ee
which are precisely the same as asymptotics of (\ref{w1}) and (\ref{w2})
respectively. Thus, the solutions to (\ref{q1}) should be identified with
the corresponding BA functions $w_{n}(z)$ and $\ov{w}_{n}(z)$:
\be  \label{w-p1}
w_{n}(z) \equiv {\cal P}^{(1)}_{n}(z)
\ee
\be \label{w-p2}
\ov{w}_{n}(z) \equiv {\cal P}^{(2)}_{n}(z)
\ee
Similarly, one can
consider the spectral problem (\ref{adLax}), (\ref{RTC1}), (\ref{RTC2}).
Then, the
corresponding $w_{n}^{\star}(z)$ and $\ov{w}_{n}^{\;\star}(z)$ satisfy
the same three-term recurrent equation as ${\cal P}^{\star}_{n}(z^{-1})
\equiv \Phi^{\star}_{n}(z^{-1})/h_{n}$ (see (\ref{PP}) and
(\ref{q2})), i.e.
\be\label{P-bar}
\frac{h_{n+1}}{h_{n}}{\cal P}^{\star}_{n+1} -
\frac{S^{\star}_{n}}{S^{\star}_{n-1}}{\cal P}^{\star}_{n} =
z^{-1} \left\{ {\cal P}^{\star}_{n} -
\frac{S^{\star}_{n}}{S^{\star}_{n-1}}{\cal P}^{\star}_{n-1}\right\}
\ee
This equation obviously has two independent solutions
with asymptotics
\be  \label{p1st}
{\cal P}^{\star (1)}_{n}(z^{-1}) = z^{-n}\Bigl(1 +
O\Bigl(\frac{1}{z}\Bigr)\Bigr) \;\;\;\;\;z\to\infty
\ee
\be \label {p2st}
{\cal P}^{\star (2)}_{n}(z^{-1}) = \frac{1}{h_{n}} z^{-n}\Bigl(1 +
 O(z)\Bigr)\;\;\;\;\;z\to 0
\ee
which are exactly of the form (\ref{ad1})
and (\ref{ad2}). Thus, the adjoint BA functions should be identified with
solutions to (\ref{P-bar}) as follows:
\be \label{w-p1st}
w_n^{\star}(z) = z^{-1}{\cal P}^{\star (1)}_{n-1}(z^{-1})
\ee
\be \label{w-p2st}
\ov{w}_n^{\;\star}(z) = z^{-1}{\cal P}^{\star (2)}_{n-1}(z^{-1})
\ee

To conclude, we see that equation
(\ref{Lax2DTL}) (see also (\ref{OE1}) below) for ${\cal P}_n(z)$,
${\cal P}^{\star}_n(z^{-1})$
is exactly the same as that for $w_{n+1}(z)$, $\ov{w}_{n+1}^{\;\star}(z)$
(compare with (\ref{OE4}), (\ref{OE5})).

\subsection{Evolution and orthogonal polynomials}

Let us now demonstrate how one can obtain evolution of the RTC using the
technique of the orthogonal polynomials.
Indeed, differentiating the orthogonality conditions with
respect to arbitrary times, one can obtain with the help of (\ref{Lax2DTL})
the evolution of polynomials  ${\cal P}_n$ and  ${\cal P}^{\star}_n$
(for details see Appendix A):
\be\label{OE1}
{\partial {\cal P}_n\over \partial t_m} =
- [({\cal L}^m)_-]_{nk}{\cal P}_k\\
{\partial {\cal P}_n\over \partial t_{-m}} = [(\ov{\cal L}^{\;m})_-]_{nk}
{\cal P}_k\\
{\partial {\cal P}^{\star}_n\over \partial t_m} = -
[({\cal L}^m)_+]_{kn}{\cal P}^{\star}_k\\
{\partial {\cal P}^{\star}_n\over \partial t_{-m}} =
[(\ov{\cal L}^{\;m})_+]_{kn}{\cal P}^{\star}_k\\
{\partial h_n\over \partial t_m} = ({\cal L}^m)_{nn}h_n\ \;\;,\;\;\;\;\;\;
{\partial h_n\over \partial t_{-m}} = - (\ov{\cal L}^{\;m})_{nn}h_n
\ee
where by definition for given matrix  $C_{nk}$,
\be
[(C)_+]_{nk} \equiv  C_{nk}\theta (k-n)\;\; ,\ \ \
(C_-)_{nk} \equiv  C_{nk}\theta (n-k-1)
\ee
i.e. $(C)_{+}$ is the upper triangular part of $C$
while $(C)_{-}$ is strictly the lower triangular part
of  $C$ .

Let us note that this evolution (of the BA functions) has
the same form as (\ref{OE4}), (\ref{OE5}). However,
the latter one was determined on the infinite
matrices, while the evolution (\ref{OE1}) -- on the quarter part of these
matrices
(this is the forced hierarchy, we will return to this point later).
In order to get the
hierarchy of non-linear equations, one should use the compatibility
conditions of the linear systems. The hierarchies obtained in this way
from the two considered evolutions do not coincide
exactly although differing only by inessential constants. Now we consider in
more details the evolution with respect to the first times to observe this
phenomenon.

Using the general formulas (\ref{OE4}), (\ref{OE5}) one gets the
simplest evolution equations
\be\label{qt11}
\frac{\d \Phi_{n}}{\d t_{1}}= S_{n}h_{n}
\sum_{k=-\infty}^{n-1}\frac{S^{\star}_{k-1}}{h_{k}}\Phi_{k}\equiv\\
\equiv  \frac{S_{n}}{S_{n-1}}\frac{h_{n}}{h_{n-1}}
(\Phi_{n} -z \Phi_{n-1})
\ee
\be \label{qt22}
\frac{\d \Phi_{n}}{\d t_{-1}} = \frac{h_{n}}{h_{n-1}} \Phi_{n-1}
\equiv (1-S_{n-1}S^{\star}_{n-1})\Phi_{n-1}
\ee
\be \label{barqt11}
\frac{\d \Phi^{\star}_{n}}{\d t_{1}} =
- \frac{h_{n}}{h_{n-1}} \Phi^{\star}_{n-1}  \equiv
-(1-S_{n-1}S^{\star}_{n-1})\Phi^{\star}_{n-1}
\ee
\be\label{barqt22}
\frac{\d \Phi^{\star}_{n}}{\d t_{-1}}
=  - S^{\star}_{n}h_{n}\sum_{k=-\infty}^{n-1}\frac{S_{k-1}}{h_{k}}
\Phi^{\star}_{k}\equiv\\
\equiv  - \frac{S^{\star}_{n}}{S^{\star}_{n-1}}
\frac{h_{n}}{h_{n-1}}(\Phi^{\star}_{n} -z^{-1}\Phi^{\star}_{n-1})
\ee

The compatibility conditions give the following nonlinear equations:
\be  \label{st11}
\frac{\d S_{n}}{\d t_{1}} = S_{n+1}\frac{h_{n+1}}{h_{n}} +
\alpha S_{n}
\ee
\be  \label{st22}
\frac{\d S_{n}}{\d t_{-1}} = S_{n-1}\frac{h_{n+1}}{h_{n}} +
\beta S_{n}
\ee
\be  \label{barst11}
\frac{\d S^{\star}_{n}}{\d t_{1}} = - S^{\star}_{n-1}\frac{h_{n+1}}{h_{n}}
-\alpha S^{\star}_{n}
\ee
\be  \label{barst22}
\frac{\d S^{\star}_{n}}{\d t_{-1}} = - S^{\star}_{n+1}\frac{h_{n+1}}{h_{n}}
-\beta S^{\star}_{n}
\ee
\be  \label{ht11}
\frac{\d h_{n}}{\d t_{1}} = - (S_{n}S^{\star}_{n-1} + \gamma)h_{n}
\ee
\be \label{ht22}
\frac{\d h_{n}}{\d t_{-1}} = (S_{n-1}S^{\star}_{n} + \gamma)h_{n}
\ee
{}From the orthogonal polynomials, one obtains the same
equations with $\alpha=\beta=\gamma=0$ \cite{KM} (see also Appendix A).
This difference, however,
does not look crucial by the following reason. One can use
identification (\ref{Shqp}) and evolutions (\ref{st1}), (\ref{barst1}) and
(\ref{ht1}) in order to get the RTC equations (\ref{q-eq2}) and
(\ref{p-eq2}) with $\nu(\epsilon)=-1$, $\epsilon=i$ and $\gamma=0$
(one can equally compare $t_{-1}$-evolutions)\footnote{The standard form
of the RTC with an arbitrary value of $\epsilon$ and $\nu$ as in (\ref{a}) can
be reached by the proper redefinition of time.}.
Let us note that the RTC itself
already does not depend on $\alpha$ and $\beta$. Therefore, besides zero
$\gamma$, one can also put $\alpha$ and $\beta$ to be zero. Put it
differently, different $\alpha$, $\beta$ and $\gamma$ describes different
representations of the same RTC hierarchy.

\section{Forced RTC hierarchy}
\subsection{RTC-reduction of 2DTL}
Now we are going to formulate in some invariant terms what reduction of
the 2DTL
corresponds to the RTC hierarchy. For doing this, let us return again to the
Lax representation (\ref{Lax2DTL}) embedding the RTC into the 2DTL.
Using (\ref{hS-rel}), one can easily prove the following identities
\be\label{zm}
\sum_{k=n}^{N}{S_{k-1}S^{\star}_{k-1}\over h_k}={\f h_N}-{\f h_{N-1}}\\
\sum_{k=n}^N S_kS^{\star}_kh_k=h_n-h_{N+1}
\ee
Because of these identities, the matrices ${\cal L}$ and ${\bar {\cal L}}^{T}$
have zero modes $\sim S_{k-1}$ and $S^{\star}_{k-1}/h_k$ respectively.
Therefore, one could naively expect that they are not invertible and get
(using (\ref{zm})) that
\be\label{LM=1}
({\cal L}\bar {\cal L})_{nk}=\delta_{nk}-{S_nS^{\star}_kh_n\over h_{-\infty}}\\
(\bar {\cal L}{\cal L})_{nk}=\delta_{nk}-{S_{n-1}S^{\star}_{k-1}h_{\infty}\over
h_k}
\ee
Since the reduction is to be described as an invariant condition imposed on
${\cal L}$ and $\bar {\cal L}$,
these formulas might serve as a starting point to describe the reduction of the
2DTL to the RTC hierarchy only if their r.h.s. does not depend on the dynamical
variables. It seems not to be the case.

However, these formulas require some careful treatment.
Indeed, the formulation of the 2DTL in terms of infinite-dimensional matrices,
although being correct as a formal construction requires some accuracy if one
wants to work with the genuine matrices since the products of the
infinite-dimensional matrices should be properly defined. In fact, this product
exists for the "band" matrices, i.e. those with only a finite number of
the non-zero diagonals, and in some other more complicated cases (of special
divergency conditions). One can easily see from  (\ref{Lax2DTL}) that the RTC
Lax operators do not belong to this class. Therefore, equations (\ref{LM=1})
just do not make sense in this case (this is why the interpretation of the
general RTC hierarchy in invariant (say, Grassmannian) terms
is a little bit complicated, see the discussion in
the last section).

Moreover, the presence of the zero mode does not mean automatically that
the matrix is not invertible since this zero mode is to be normalizable
(this is the counterpart of the band structure of matrices, the number
of non-zero entries in the column describing the zero mode is to be finite,
or to satisfy some fast-decreasing conditions).\footnote{It is evident from
the following simple example. Let us consider the
band matrix $B\equiv I-\Lambda$. This matrix is invertible, but its
inverse matrix $I+\Lambda+\Lambda^2+\ldots$ does not possess the band
structure.
At the same time, there exists the zero mode $f_j=
\hbox{const}$ of the matrix $B_{ij}$,
but it is non-normalizable since the product
$f^T f=\left(\sum_{j=-\infty}^{+\infty} \hbox{const}\right)$
is divergent.}

However, in the case of forced hierarchy, some of the indicated problems are
removed since one needs to multiply only
quarter-infinite matrices and, say, the product ${\cal L}\bar{\cal L}$
always exists. Certainly the inverse order of the multipliers is still
impossible. Therefore, only the first formula in (\ref{LM=1})
becomes well-defined acquiring the form
\be\label{forcedLM=1}
({\cal L}\bar {\cal L})_{nk}=\delta_{nk}
\ee
This formula can be already
taken as a definition of the RTC-reduction of the 2DTL
in the forced case as it does not depend on dynamical variables.
Now we will show how this definition is reflected
in different formulations of the 2DTL.

\subsection{Fermionic representation}

Here we would like to describe the 2DTL
hierarchy in terms of the massless fermions \cite{DJKM} and to describe
manifestly in the subspace in the Grassmannian corresponding to the
RTC-reduction. In this and the next subsections we closely follows
the papers \cite{KMMOZ} and \cite{KMMM} where further technical details
can be found.

Let us define the fermionic operators on sphere
\beq
\psi (z) =\sum _{k\in {\bf Z}}\psi _kz^k\hbox{ , }  \psi ^\ast (z)
=\sum _{k\in {\bf Z}}\psi ^\ast _kz^{-k}
\eeq
with fermionic modes satisfying the usual anti-commutation relations:
\beq\label{A2}
\{\psi _k\hbox{, } \psi ^\ast _m\} = \delta _{km}\hbox{ , }  \{\psi _k\hbox{, }
\psi _m\} = \{\psi ^\ast _k\hbox{, } \psi ^\ast _m\} = 0
\eeq
The Dirac vacuum  $|0\rangle $  is defined by the conditions:
\beq
\psi _k|0\rangle  = 0\ ,\ k < 0\ ;\ \psi ^\ast _k|0\rangle  = 0\ ,\ k \geq
0
\eeq
We also need to introduce the ``shifted" vacua
\beq
\psi _k|n\rangle  = 0\ ,\ k < n\hbox{ ; }   \psi ^\ast _k|n\rangle  = 0\ ,\ k
\geq  n
\eeq
{}From fermionic modes one can built the  $U(1)$--currents
\beq
J_k =\sum _{i\in {\bf Z}}\psi _i\psi ^\ast _{i+k}\hbox{ , }  J_{-k} \equiv
\bar J_k\hbox{ , }  k \in  {\bf Z}_+
\eeq
and define ``Hamiltonians"
\beq
H(t_{k}) = \sum ^\infty _{k=1}t_kJ_k\hbox{ , }  \bar H(t_{-k}) =
\sum ^\infty _{k=1}t_{-k}\bar J_k
\eeq
where  $\{t_k\}$  and  $\{t_{-k}\}$  are ``positive" and ``negative" times
correspondingly, which
generate the evolution of nonlinear system.

Let  $g$  be an arbitrary element of the Clifford group which does not mix
the $\psi$- and $\psi ^\ast $- modes :
\beq\label{A8}
g = :\exp \left[\sum     A_{km}\psi _k\psi ^\ast _m\right]:
\eeq
where  :  :  denotes the normal ordering with respect to the Dirac vacuum
$|0\rangle $. Then it is well known (see \cite{KMMM} and references therein)
that
\beq\label{A9}
\tau _n(t) = \langle n|e^{H(t_{k})}ge^{-\bar H(t_{-k})}|n\rangle
\eeq
solves the two-dimensional Toda lattice hierarchy, i.e. is the solution to
the whole set of the Hirota bilinear equations. Any particular solution depends
only on the choice of the element  $g$ (or, equivalently, it can be uniquely
described by the matrix  $A_{km}$). From relations (\ref{A2})
one can conclude that
any element of the form (\ref{A8}) rotates the fermionic modes as follows
\beq\label{R}
g\psi _kg^{-1} = \psi _jR_{jk}\hbox{ , }  g\psi ^\ast _kg^{-1} =
\psi ^\ast _jR^{-1}_{kj}
\eeq
where the matrix  $R_{jk} $ can be expressed through  $A_{jk}$. We
will see below that the general solution (\ref{A9}) can be expressed in the
determinant form with explicit dependence on  $R_{jk}$ .

Now we deal with the forced hierarchies, i.e. impose the condition
\beq\label{forced}
\tau _n = 0\hbox{  , }  n < 0
\eeq
Let us determine what substitutes the general expression (\ref{A8}) in this
case.
It is reasonable to look for the point of the Grassmannian in the
form
\beq\label{B1}
g = g_0P_+
\eeq
where  $P_+$ is the projector onto positive states:
\beq
P_+|n\rangle  = \theta (n)|n\rangle
\eeq
It can be constructed from the fermionic modes as follows
\beq
P_+ = :\exp \left(\sum _{i<0}\psi _i\psi ^\ast _i\right):
\eeq
and enjoy the properties
\be
P_+\psi ^\ast _{-k} = \psi _{-k}P_+ = 0\hbox{  , }  k > 0\ ;\
\left[ P_+,\psi _k\right] = [P_+,\psi ^{\ast} _k] = 0\hbox{  , } k \geq  0\ ;\
P^2_+ = P_+
\ee

The insertion of such a projector into eq.(\ref{A9}) leaves us with $g_0$
depending only on $\psi _k$ and $\psi ^\ast _k$ with $k \geq
0$. Therefore, it has the form \cite{KMMM}
\beq\label{B2}
g_0 = :\exp \left\{ \left(\int _\gamma  A(z,w)\psi _+(z)
\psi ^\ast _+(w^{-1})dzdw \right) - \sum _{i\geq 0} \psi _i\psi ^\ast _i
\right\}:
\eeq
where $\psi _+(z) =\sum _{k\geq 0}\psi _kz^k$ , $\psi ^\ast _+(z)
=\sum _{k\geq 0}\psi ^\ast _kz^{-k}$ and  $\gamma $  is some
integration contour. The matrix elements $A_{ij}$ from (\ref{A8}) are
immediately connected with
the modes of the function $A(z,w)$ and, in the forced case, this matrix
coincides with the quarter of the matrix $R_{ij}$ from (\ref{R})
(the rest part of $R_{ij}$ is just unit). Now we are ready to describe
manifestly the RTC reduction of the forced hierarchy. The reduction condition
(\ref{forcedLM=1}) means that the matrix $A_{ij}$ celebrates the property
\be\label{A}
\Lambda_+\cdot A\cdot\Lambda_-=A
\ee
where $\Lambda_+$ and $\Lambda_-$ are the quarter-infinite matrices,
$(\Lambda_+)_{ij}\equiv \delta_{i,j-1}$,
$(\Lambda_-)_{ij}\equiv \delta_{i-1,j}$. These matrices are not invertible
(they would be inverse to each other in the case of infinite matrices),
although $\Lambda_+\cdot\Lambda_-=1$. The point is that the inverse order
in this product does not lead to the unit matrix. By the same reason,
condition (\ref{A}) can not be replaced, say, by the similar condition
$\Lambda_+A=A\Lambda_+$ (or any of remaining two possibilities).
This reflects the fact that only the first of
eqs.(\ref{LM=1}) makes good sense in the forced RTC case.

Condition (\ref{A}) means that $A_{ij}=A_{i-j}$ and
\be\label{B}
A(z,w)={\mu'(z)\over 2\pi iz}\delta(z-w^{-1})
\ee
where $\mu(z)$ is some arbitrary measure (compare with (\ref{sp})).

Thus, RTC-reduction is described by Toeplitz matrices
$A_{ij}=A_{i-j}$ which give rise to the
corresponding subspace in the whole Grassmannian given by arbitrary matrices
$A_{ij}$. In the next subsection, we demonstrate this by the direct calculation
from the unitary matrix model representation of the RTC.

\subsection{Determinant representation}

One can show from formulas (\ref{B1})-(\ref{B2}) that $\tau$-function
(\ref{A9})
of the forced hierarchy has the
following determinant representation \cite{KMMM}
\beq\label{detrep}
\tau _n(t) = \left.
\det \left[\partial ^i_{t_1}(-\partial _{t_{-1}})^j\int _\gamma
A(z,w)\exp\left\{\sum_{m>0} (t_mz^m- t_{-m}w^{-m})\right\}dzdw\right]
\right|
_{i,j=0,...,n-1}
\eeq
Let us demonstrate now that the unitary matrix model (= forced
RTC) leads to this
determinant representation with $A$ as in (\ref{B}).

We begin with rewriting the orthogonality relation (\ref{orth-cond})
in the matrix form. For doing this, we define matrices $D$ and $D^{\star}$
with the matrix elements determined as the coefficients of the polynomials
$\Phi_n(z)$ and $\Phi_n^{\star}(z)$
\be
\Phi_i(z)\equiv \sum_jD_{ij}z^{j-1},\ \ \ \Phi^{\star}_i(z)\equiv
\sum_j D^{\star}_{ij}z^{j-1}
\ee
Then, (\ref{orth-cond}) looks like
\be\label{oco}
D\cdot C\cdot D^{\star T}=H
\ee
where superscript $T$ means transponed matrix and $H$ denotes the diagonal
matrix with the entries $C_{ii}=h_{i-1}$
and $C$ is the moment matrix with the matrix
elements
\be
C_{ij}\equiv \int_{\gamma} {d\mu(z)\over 2\pi iz} z^{i-j}
\exp\left\{\sum_{m>0}(t_mz^m-t_{-m}z^{-m})
\right\}
\ee
Let us note that $D$ ($D^{\star T}$) is
the upper (lower) triangle matrix with the units on the diagonal (because of
(\ref{polyn})). Indeed, this representation is nothing but the
Riemann-Hilbert problem for the forced hierarchy.
Now taking the determinant of the both sides of (\ref{oco}),
one gets
\be
\det_{n\times n} C_{ij}=\prod_{k=0}^{n-1} h_k =\tau_n
\ee
due to formula (\ref{prod}). The remaining last step is to observe that
\be\label{moma}
C_{ij}=\partial ^i_{t_1}(-\partial _{t_{-1}})^j\int _\gamma
{d\mu(z)\over 2\pi iz}
\exp\left\{\sum_{m>0}(t_mz^m-t_{-m}z^{-m})\right\}=
\partial ^i_{t_1}(-\partial _{t_{-1}})^j C_{11}
\ee
i.e.
\be\label{taumoma}
\tau_n(t)=\det_{n\times n}\left[
\partial ^i_{t_1}(-\partial _{t_{-1}})^j\int _\gamma
{d\mu(z)\over 2\pi iz}
\exp\left\{\sum_{m>0}(t_mz^m-t_{-m}z^{-m})\right\}
\right]
\ee
This expression coincides with (\ref{detrep}) with $A(z,w)$ chosen as in
(\ref{B}). One can also remark that the moment matrix $C_{ij}$ is Toeplitz
matrix satisfying formula (\ref{A}). This proves from the different
approach that the RTC-reduction is defined by the Toeplitz matrices.

\section{Relativistic Toda molecule}

\subsection{General properties}

In this section we consider further restrictions on the RTC which allows one
to consider the {\it both} products in (\ref{LM=1}). Namely, in addition to
the condition (\ref{forced}) picking up forced hierarchy, we impose the
following constraint
\be\label{molecule}
\tau_n=0,\ \ \ n>N
\ee
for some $N$.
This system should be called $N-1$-particle
relativistic Toda molecule, by analogy
with the non-relativistic case and is nothing but RTC-reduction of the
two-dimensional Toda molecule \cite{TM,LS}\footnote{Sometimes
the Toda molecule
is called non-periodic Toda \cite{OP}. It is an
immediate generalization
of the Liouville system.}.

Now we describe this system in different representations. Let us start with
the general two-dimensional Toda molecule and discuss which
element $g$ in (\ref{A8}) describes this hierarchy.
Since the Toda molecule is the
very particular case of the forced hierarchy, we can look at representation
(\ref{B2}). Then, the $sl(N)$ Toda can be described by the matrix $A_{ij}$
of finite rank $N$ \cite{QTF2}.
This means that it can be presented as the finite sum
\be\label{ATM}
A_{ij}=\sum_k^N f^{(k)}_ig^{(k)}_j
\ee
where $f^{(k)}_i$ and $g^{(k)}_j$ are arbitrary coefficients. For the kernel
$A(z,w)$ (\ref{B2}), this condition is
\be\label{kernel}
A(z,w)=\sum_k^N f^{(k)}(z)g^{(k)}(w)
\ee
where $f^{(k)}(z)$ and $g^{(k)}(z)$ are arbitrary functions.
Indeed, these functions (or the sets of coefficients) describe the
way how $sl(N)$ group can be embedded into $gl(\infty)$ (different embeddings
are
related by the external $gl(\infty)$ automorphisms of $gl(N)$) \cite{QTF2} --
this is why the system is sometimes called $sl(N)$ Toda molecule.
{}From this description, one can immediately read off the corresponding
determinant representation (\ref{detrep}).

Indeed, equation (\ref{todaeq}) and condition (\ref{molecule}) implies
that $\log\tau_0$ and $\log\tau_N$ satisfy the free wave equation
\be\label{wave}
\partial_{ t_1}\partial_{ t_{-1}}\log \tau_0=
 \partial_{ t_1}\partial_{ t_{-1}}\log \tau_{N}=0
\ee
Since the relative normalization of $\tau_n$'s is not fixed, we are free
to choose $\tau_0=1$. Then,
\be \label{mol}\tau_{0}(t)=1\;\;,\;\;\;\;\;
	\tau_{N}(t) = \chi(t_1)\bar \chi( t_{-1})
\ee
where $\chi(t_1)$ and $\bar \chi(t_{-1})$ are arbitrary functions. 2DTL with
boundary conditions (\ref{mol}) was considered in \cite{TM}. The solution
to (\ref{todaeq}) in this case is given by \cite{LS}:
\be\label{C1}
\tau_{n}(t)\;=\;
	\det\;\partial_{t_1}^{i-1} (-\partial_{t_{-1}})^{j-1}\tau_{1}(t)
\ee
with
\be
\tau_{1}(t) = \sum_{k=1}^{N} a^{(k)}(t)\bar a^{(k)}(t_{-1})
\ee
where functions $a^{(k)}(t)$ and $\bar a^{(k)}(t_{-1})$ satisfy
\be
\det\partial^{i-1}_{t_1}a^{(k)}(t)=\chi(t)\; , \;\;
	  \det (-\partial_{t_{-1}})^{i-1}\bar a^{(k)}(t_{-1})=\bar \chi(t_{-1})
\ee

This result coincides with that obtained by substituting into (\ref{detrep})
the kernel $A(z,w)$ of the form (\ref{kernel}).

Let us stress that, although the Toda molecule looks as the forced hierarchy
with one more projector inserted, this is described by the infinite number
of the fermionic modes, since infinitely many matrix elements of (\ref{ATM})
are not zero. However, the Toda molecule can be described by the finite matrix
Lax representation how we shall see in the next subsection.

Now we have to describe the RTC-reduction of the Toda molecule. It is
clear that, to do this, one needs to impose on the $C_{11}$-element of
the moment matrix (\ref{moma}) to have the structure (\ref{C1}), or, which is
the same, to require that the matrix
$\partial_{t_1}^{i-1} (-\partial_{t_{-1}})^{j-1}
\sum_{k=1}^{N} a^{(k)}(t)\bar a^{(k)}(t_{-1})$ to be Toeplitz. These
conditions leads to some equations with the solution just giving the RTC
Toda molecule, i.e. picking up the RTC-reduction
among all the Toda molecule solutions. The same reduction also can be
formulated as the condition for matrix (\ref{ATM}) to be Toeplitz which
looks quite tricky since naively leads to overfulled system of equations.

However, in the last subsection of this section we investigate the simplest
$sl(2)$
case in detail and demonstrate that these conditions have the non-trivial
solutions, i.e. the relativistic Toda molecule does really exist. Indeed,
how is already clear from the next subsection, the RTC Toda molecule
is the system with finite ($N-1$) degrees of freedom, described by
$N-1$ independent time flows and possessing, therefore, finitely many
conservation laws. Unlike this, the whole 2DTL hierarchy, and its any
reductions considered above (RTC, forced etc) are the systems with infinitely
many degrees of freedom. The RTC molecule is so "small" because of rigidity
of the two simultaneous reductions: the RTC and Toda molecule ones. In a sense,
these two reductions are almost "perpendicular" remaining very small room
for common solutions.

\subsection{Lax representation}

In all our previous considerations, we dealt with infinite-dimensional
matrices.
Let us note that the Toda molecule can be effectively treated in terms of
$N\times N$ matrices like the forced case could be described by the
quarter-infinite matrices. This allows one to deal with the {\it both}
identities
(\ref{LM=1}) since all the products of {\it finite} matrices are well-defined.

To see this, one can just look at the recurrent relation (\ref{Lax2DTL}) and
observe that there exists
the finite-dimensional subsystem of ($N$) polynomials which is decoupled from
the whole system. The recurrent relation for these polynomials can be
considered
as the finite-matrix Lax operator (which still
does not depend on the spectral parameter, in contrast to (\ref{l-op})).
Indeed, from (\ref{Lax2DTL}) and
condition (\ref{molecule}), i.e. $h_n/h_{n-1}=0$ as $n\ge N$ (the Toda
molecule conditions in terms of $S$-variables read as $S_n=S^{\star}_n=1$
for $n>N-2$ or $n>0$),
one can see that
\be
z{\cal P}_N(z)={\cal P}_{N+1}(z)-{\cal P}_N(z),\ \ \
z{\cal P}_{N+1}(z)={\cal P}_{N+2}(z)-{\cal P}_{N+1}(z)\ \ \ \hbox{etc.}
\ee
i.e. all the polynomials ${\cal P}_n$ with $n>N$ are trivially expressed
through ${\cal P}_N$. Therefore, the system can be effectively described by
the dynamics of only some first polynomials (i.e. has really finite number of
degrees of freedom). Certainly, all the same is correct for the
star-polynomials ${\cal P}_n^{\star}$ although, in this case, it would be
better
to use the original non-singularly normalized polynomials $\Phi_n^{\star}$.

Now let us look at the corresponding Lax operators (\ref{RTC1})-(\ref{RTC2}).
They are getting quite trivial everywhere but in the left upper corner
of the size $N\times N$. For instance,
\be\label{L1}
{\cal L}=\left(
\begin{array}{cccccccc}-S_0 & 1 &&&\vdots&&&\\
-{h_1\over h_0}S_1& -S_1S^{\star}_0&1&&\vdots&&&\\
&  \ldots & &1&\vdots&&&\\
\ldots & -{h_{N-1}\over h_k} S^{\star}_{k-1} &\ldots&
-S^{\star}_{N-2}&\vdots &&&\\
\multicolumn{4}{c}\dotfill&-1&1&&\\
&&&&&-1&1&\\
&&&&&&-1&1\\
&&&&&&&\ddots
\end{array}
\right)
\ee
and analogously for the Lax operator $\bar {\cal L}$. Therefore, one can
restrict himself to the system of $N$ polynomials ${\cal P}_n$, $n=0,1,\ldots,
N-1$ and the finite matrix Lax operators (of the size $N\times N$).

Now one needs only to check that this finite system still has the same
evolution equations (\ref{OE1}). It turns out to be the case only for the
first $N-1$ times. This is not so surprising, since, in the finite system
with $N-1$ degrees of freedom, only first $N-1$ time flows are independent.
Therefore, if looking at the finite matrix Lax operators, one gets the
dependent higher flows. On the other hand, if one embeds this finite system
into
the infinite 2DTL, one observes that the higher flows can be no longer
described inside this finite system. Let us remark that just the finite
system is often called relativistic Toda molecule (see, for example,
\cite{FM}).)

To simplify further notations, we introduce, instead of $S_n$, $S^{\star}_n$,
the new variables $s_n\equiv (-)^{n+1} S_n$, $s^{\star}_n\equiv (-)^{n+1}
S^{\star}_n$. Then, one can realize
a very interesting property of the Lax operator (\ref{L1})
-- it can be constructed as the product of simpler ones:
\be\label{L2}
{\cal L}={\cal L}_N{\cal L}_{N-1}\ldots{\cal L}_1
\ee
where ${\cal L}_k$ is the unit matrix wherever but a $2\times 2$-block:
\be\label{L4}
{\cal L}_k\equiv
\left(\begin{array}{ccc}
1&\vdots&\\
\cdots&G_k&\cdots\\
&\vdots&1
\end{array}
\right)\ \ \ \ \ \ G_k\equiv\left(\begin{array}{cc}
s_k&1\\
s_ks^{\star}_k-1&s^{\star}_k
\end{array}\right)
\ee
Analogously
\be\label{L3}
\bar {\cal L}=\bar {\cal L}_1\ldots\bar {\cal L}_{N-1}\bar {\cal L}_N
\ee
with
\be
\bar {\cal L}_k\equiv
\left(\begin{array}{ccc}
1&\vdots&\\
\cdots&\bar G_k&\cdots\\
&\vdots&1
\end{array}
\right)\ \ \ \ \ \ \bar G_k\equiv\left(\begin{array}{cc}
s_k^{\star}&-1\\
1-s_ks^{\star}_k&s_k
\end{array}\right)
\ee
One can trivially see that ${\cal L}_k\bar {\cal L}_k=\bar {\cal L}_k
{\cal L}_k=1$, and, therefore, from (\ref{L2}) and (\ref{L3}), one obtains
${\cal L}\bar {\cal L}=\bar {\cal L} {\cal L}=1$ (cf. (\ref{LM=1})).

{}From formulas (\ref{L2})-(\ref{L3}), one trivially obtains that
$\det {\cal L}=\det \bar {\cal L}=1$ which reminds once more of the $sl(N)$
algebra. More generally, the factorization property of the
Lax operators opens the wide road to the
group theory interpretation of the RTC molecule. Indeed, following the
line of papers \cite{QTF1,QTF2,FM}, one should identify the (family of)
solutions to the integrable hierarchies with representations of
the algebra of functions on the underlying group manifold.
In the quantum group case, one should
expect that different reductions correspond to fixing the irreducible
representations \cite{QTF1,QTF2}. As the classical counterpart of this
statement,
one should consider the (quadratic) Poisson structure for the group
elements\footnote{In paper \cite{RST}, the same ideology was applied to
the linear Poisson structures. This leads to the
non-relativistic TC, in accordance with the "algebraic" character of this
latter
as opposed to the "group" character of the RTC.}
\be\label{Poisson}
{}\{g \stackreb{{{,}\atop{\phantom{g}}}}{\otimes}g\}=[r,g\otimes g]
\ee
where $r$ is the classical $r$-matrix (see, e.g., \cite{FT}),
and, instead of irreducible representations,
look at the simplectic leaves, i.e. the submanifolds where
Poisson structure is non-degenerate (an analog of the geometric quantization)
-- see \cite{Soi}. This procedure has been performed
in \cite{FM} for the $sl(N)$
group and was shown to lead to the relativistic Toda molecule for
the leaves of the dimension equal to the doubled rank of the group
(i.e. $2(N-1)$ for $sl(N)$).

It was shown in \cite{Soi} that there always exist such leaves\footnote{When
quantizing, they lead to the infinite-dimensional representations of the
algebra
of functions.} and they were manifestly described. Namely, for the $sl(2)$
case,
the 2-dimensional
simplectic leaf can be described by the group element of the form
\be
g=\left(\begin{array}{cc}
a&b\\b&d
\end{array}\right)\ \ \ \ \ ad-b^2=1
\ee
Multiplying this matrix by the trivial rescaling matrix
$
g=\left(\begin{array}{cc}
b^{-1}&\\&b
\end{array}\right)
$
one can easily transform $g$ to the matrix $G_k$ in (\ref{L4}).

In fact, the only important information of this matrix
at the moment is that it is parametrized by a 2-dimensional manifold.
However, the crucial test comes when considering the structure of this
matrix for higher rank groups. Namely,
for arbitrary $sl(N)$ group, the group element corresponding to the
proper simplectic leaf is constructed from elementary $sl(2)$ building
blocks exactly how it is done in formula (\ref{L2}) \cite{Soi,FM}.
Therefore, the Lax
operator of the relativistic $sl(N)$ Toda molecule can be treated as the group
element of the special form corresponding to the simplectic leaf of the
dimension $2(N-1)$. Then, one can construct the Hamiltonians commuting with
respect to the Poisson structure (\ref{Poisson}) and giving the time flows
just as traces of the Lax operator.

It may seem that there does not remain any room for the other Lax operator
(\ref{L3}). However, this is not the case. Indeed, the above construction
is correct up to any Weyl transformation. The transformation corresponding to
the longest element of the Weyl group maps ${\cal L}$ operator to
$\bar {\cal L}$ operator.
This explains the symmetry of dynamics with respect to the positive and
negative times. As for other Weyl transformations, the corresponding
Hamiltonians can be hardly  into 2DTL dynamics.

\subsection{$sl(2)$ molecule}

To illustrate the results of this section, let us consider the simplest $sl(2)$
example of the relativistic Toda molecule. This case is to be considered as a
"relativization" of the Liouville theory. Surprisingly enough, this case turns
out to be {\it equivalent} to the Liouville theory. This is a specific
feature of the $sl(2)$ case, which is no longer correct already for the $sl(3)$
group. This can be demonstrated by comparing the Lax operators, but we use
more immediate way of comparing solutions to the equations of hierarchy.

In fact, the $sl(2)$ hierarchy corresponding to the rank one group
contains only two equations: one for positive and the other
one for negative times. One of these equations is (\ref{rt}) with the
Toda molecule constraints $q_{-1}\to \infty$, $q_2\to -\infty$ etc.
Under these constraints, there are only two nontrivial RTC equations
\be\label{E1}
\ddot q_0 = (1+\epsilon\dot q_0)(1+\epsilon\dot q_1){e^{q_1-q_0}\over
1+\epsilon^2e^{q_1-q_0}}\\
\ddot q_1 =-(1+\epsilon\dot q_0)(1+\epsilon \dot q_1){e^{q_1-q_0}\over
1+\epsilon^2e^{q_1-q_0}}
\ee
These equations imply that $q_0+q_1=C_1t+C_2$ with arbitrary constants
$C_1,\ C_2$. Now introducing the new variable $q\equiv q_0-{\2}(C_1t+C_2)$
one gets the equation
\be\label{E2}
\ddot q = \left[(1+\2\epsilon C_1)^2-\epsilon^2\dot q^2\right]{e^{-2q}\over
1+\epsilon^2e^{-2q}}
\ee
The limit of $\epsilon\to 0$ leads to the Liouville equation
\be
\ddot q = e^{-2q}
\ee
Equation (\ref{E2}) has the solution
\be\label{RTCsol}
\int {dq\over\sqrt{{(1+\2\epsilon C_1)^2+C_3\over \epsilon^2}+C_3e^{-2q}}}
=t_1+C_4\\ \hbox{i.e.}\ \ \ q=\log \left[A_0\sinh (A_1t_1+A_2)\right]
\ee
(with some redefined constants $\{C_i\}\to \{A_i\}$)
which is to be compared with the analogous solution to the Liouville equation
(non-relativistic Toda molecule)
\be
\int {dq\over\sqrt{C_3-e^{-2q}}}
=t_1+C_4
\ee
One easily observes that these both can be transformed to each other by the
trivial rescaling of the time variable and integration constants. Therefore,
"the relativization" of the Liouville equation does not lead to a new
equation.

Let us note, however, that switching on the negative time changes this
statement. Indeed, looking at equation completely
equivalent to (\ref{rt}), one can check that the solutions really depend
on an arbitrary linear combination of times:
\be\label{qsol2}
q=\log \left[A_0\sinh (A_1t_1+A_3t_{-1}+A_2)\right]
\ee
Therefore, the notorious property of dependence only on the sum of positive
and negative times, which distinguishes the TC hierarchy
is not fulfilled for the RTC molecules. Thus, these systems are really
different even in the $sl(2)$ case.

In fact, the pair of equations (\ref{E1}) can be really substituted by the
only equation for the $\tau$-function $\tau_1$ (\ref{todaeq})
with the Toda molecule (\ref{mol}) and reduction conditions. To do this, let
us first note that conditions (\ref{wave}) allows one to normalize
$\tau$-functions so that $\tau_0=\tau_N=1$. In particular, in the $sl(2)$
case this means that one can add to the equation (\ref{todaeq}) for
$\tau_1$ analogous equations for $\tau_0$ and $\tau_2$ with the coefficient
$-\2$ to get
\be
\2\partial_{t_1}\partial_{t_{-1}}\log {\tau_1^2\over\tau_0\tau_2}=-
{\tau_0\tau_2\over\tau_1^2}
\ee
and, for the new function $\tau\equiv {\tau_1\over\sqrt{\tau_0\tau_2}}$,
\be\label{Etau}
\partial_{t_1}\partial_{t_{-1}}\log \tau=-
{1 \over\tau^2}
\ee
Using formulas (\ref{Shqp}) and (\ref{tau}), one obtains that
\be\label{qtau}
\epsilon^2
e^{q_1-q_0}=-{\tau_0\tau_2\over\tau_1^2}=-{1\over\tau^2}=
\epsilon^2e^{-2q}
\ee
We have already remarked (see the very end of sect.3.1) that, generically,
the connection of $\tau$-functions with $q$-variables may require some
additional shift of these latter.

Now one should solve equation (\ref{Etau}) with the moment matrix
$\partial_{t_1}^{i-1} (-\partial_{t_{-1}})^{j-1}\tau_{1}(t)$
constrained to be Toeplitz. In our case, this means that
$\partial_{t_1}\partial_{t_{-1}}\tau=-\tau$. Then, one obtains the equation
(the same equation can be obtained from (\ref{E2}) by the substitute
(\ref{qtau}))
\be\label{taueq}
\tau={\tau^2-1\over(\partial_{t_1}\tau)^2}\partial_{t_1}^2\tau
\ee
This equation has the solution
\be\label{El}
\int {d\tau\over\sqrt{\tau^2+1}}
=B_1t_1+B_2
\ee
i.e.
\be\label{taumol}
\tau=\sinh (B_1t_1+B_2)
\ee
and can be reduced to (\ref{RTCsol}) through the replace (\ref{qtau})
$\tau={i\over\epsilon}e^{q+\hbox{const}}$.
This formula differs from (\ref{qtau}) by a
constant shift of $q$. Indeed, we have already remarked
(see the very end of sect.3.1) that, generically,
the connection of $\tau$-functions with $q$-variables may require some
additional shift of these latter. In the present case,
the shift depends on value of $C_1$
and is in charge of the non-unit coefficient $A_0$ in
(\ref{qsol2}). The correct connection of $\tau$ and $q$ restores if
$C_1$ is chosen to be $-{2\over\epsilon}$, i.e. $q_n\to q_n-{1\over\epsilon}
t_1$.

Now let us look at the dependence of $\tau$ on
the negative time. Naively, using (\ref{taumol})
and constraint $\partial_{t_1}\partial_{t_{-1}}\tau=-\tau$, one gets that
the general solution is given by the formula
\be
\tau=\sinh (B_1t_1-B_1^{-1}t_{-1}+B_2)
\ee
which contradicts to (\ref{qsol2}) with arbitrary unrelated coefficients
in front of times. This puzzle is solved by noting that the moment
matrix (\ref{moma}) can be multiplied by arbitrary triangle matrix
with units on the diagonal so that the determinant (\ref{taumoma}) does not
change. This means that one really needs to consider the weaker
Toeplitz condition
$\partial_{t_1}\partial_{t_{-1}}\tau+\alpha\partial_{t_{-1}}\tau=-\tau$,
where $\alpha$ is a constant (this does not effect equation (\ref{taueq})).
This results to the general solution for
$\tau$-function
\be
\tau=\sinh (B_1t_1+B_3^{-1}t_{-1}+B_2)
\ee
consistent with (\ref{qsol2}).
This provides us with some nontrivial
example of the simultaneous
solutions to the constraints of the relativistic Toda molecule
and Toeplitz moment matrix. The measure $\mu'(z)$ (\ref{B}) can be obtained
from the explicit formulas (\ref{taumol}) and (\ref{taumoma})
by the Fourier transform.

\sect{Discrete evolutions and limit to Toda chain}
\subsection{Darboux-B\"acklund transformations}
In this section, we are going to discuss some discrete evolutions of the RTC
given by the Darboux-B\"acklund transformations and their limit to the usual
Toda chain. One can easily take the continuum limit of the formulas
of this section to reproduce the TC as the limit of the RTC, both with
the standard continuous evolutions.

The discrete evolution equations in the RTC framework were recently
introduced by \cite{Sur} in a little bit sophisticated way. Here we outline the
simple approach based on the notion of the Darboux-B\"acklund
transformation (DBT). More
details will be presented in the separate publication \cite{KMZ}.

Let discrete index $\,i\,$ denote the successive DBT's.
The spectral problem now can be written as follows:
\be\label{Rsp}
\Phi_{n+1}(i|z) +a_n(i)\Phi_n(i|z) = z\Bigl\{ \Phi_n(i|z) +
b_n(i)\Phi_{n-1}(i|z)\Bigr\}
\ee
In fact, we consider two pairs of different DBT's, the two in each pair being
complimentary to each other. This is why we call the DBT's in a pair
forward and backward DBT's.

Let us define the first forward DBT (treating it as the
discrete evolution) in the form very similar
to that of the usual Toda chain:
\be \label{Rf1}
\Phi_n(i+1|z) = \Phi_n(i|z) + \balpha^{(1)}_n(i) \Phi_{n-1}(i|z)
\ee
where $\,\balpha^{(1)}_n(i)\,$ are some unknown functions.
{}From the point of view of the whole hierarchy, this new evolution means
nothing but
a discretized version of the first negative flow equation (\ref{todaev}).
One requires that $\;\Phi_n(i+1)\;$ satisfies the same spectral problem as
(\ref{Rsp})
but with the shifted value of $\,i\,$:
\be\label{Rsp2}
\Phi_{n+1}(i+1|z) +a_n(i+1)\Phi_n(i+1|z) = z\Bigl\{ \Phi_n(i+1|z) +
b_n(i+1)\Phi_{n-1}(i+1|z)\Bigr\}
\ee
Then the compatibility condition gives the equations of the discrete RTC:
\be\label{fa1}
a_n(i+1) = a_{n-1}(i)\;\frac{a_n(i)-\balpha^{(1)}_{n+1}(i)}{a_{n-1}(i)-
\balpha^{(1)}_{n}(i)}
\ee
\be\label{fb1}
b_n(i+1) = b_{n-1}(i)\;\frac{b_n(i)-\balpha^{(1)}_{n}(i)}{b_{n-1}(i)-
\balpha^{(1)}_{n-1}(i)}
\ee
\be\label{fz1}
z_i\frac{b_n(i)}{\balpha^{(1)}_n(i)} =
z_i - a_n(i) + \balpha^{(1)}_{n+1}(i)
\ee
where $\;z_i\;$ are arbitrary constants.
% (\ref{fz1}) is the direct analog of (\ref{aa});
These equations give some natural generalization of the corresponding discrete
evolution for the Toda chain hierarchy; moreover, there exists the simple limit
to the usual discrete Toda equations (see the discussion of this limit below).
Using representation (\ref{a-rep}), (\ref{b-rep}) it is very easy to rewrite
the system (\ref{fa1})-(\ref{fz1}) in terms of "coordinates" $\,q_n(i)\,$ only
\cite{Sur}.

{}From the continuum picture, one can get that there is another simple
evolution
($t_1$-flow)
treated in a parallel way with that generated by the $\,t_{-1}$-flow.
Therefore, one can try to find the corresponding discrete analog of
(\ref{todaev2}). Indeed, such analog exists and has the natural form
\be\label{Rf2}
\Phi_n(i+1|z) = \bigl(1-\balpha^{(2)}_n(i)\bigr)\Phi_n(i|z) +
z\balpha^{(2)}_n(i)\Phi_{n-1}(i|z)
\ee
where $\balpha^{(2)}_n(i)\,$ are some new unknown functions of the
corresponding discrete indices. We refer to this evolution as to the
second forward DBT.
Substitution of (\ref{Rf2}) to (\ref{Rsp}) gives quite different system of the
discrete evolution equations:
\be\label{fa2'}
a_n(i+1)=
a_n(i)\frac{1-\balpha^{(2)}_{n+1}(i)}{1-\balpha^{(2)}_{n}(i)})
\ee
\be\label{fb2'}
b_{n}(i+1) =
b_{n-1}(i)\frac{\balpha^{(2)}_n(i)}{\balpha^{(2)}_{n-1}(i)}
\ee
\be \label{fz2'}
a_n(i) + b_n(i)\;\frac{1-\balpha^{(2)}_n(i)}{\balpha^{(2)}_n(i)} =
z_i\;\frac{1}{1-\balpha^{(2)}_{n+1}(i)}
\ee
This system is the discrete counterpart of the continuum system (\ref{kuz1}),
(\ref{kuz2}).
Written in the terms of $\,q_n(i)\,$, equations (\ref{fa2'})-(\ref{fz2'})
define the same evolution of coordinates as in the case of the first
forward DBT.

Actually, in \cite{Sur}, four different discrete systems of the RTC equations
were written.
{}From our point of view, the additional evolutional systems result from the
{\bf
backward Darboux-B\"acklund transformations} which are complimentary to those
described
above. For example, the first backward DBT has the form
\be\label{Rb1}
\Phi_n(i|z) = \frac{1}{z-z_i}
\Bigl\{\bigl(1-\bbeta^{(1)}_n(i)\bigr)\Phi_{n+1}(i+1|z) +
z\bbeta^{(1)}_n(i)\Phi_n(i+1|z)\Bigr\}
\ee
where $\;z_i\;$ are the same constants as in (\ref{fz1}) and variables
$\;\bbeta^{(1)}_n(i)\;$  are related with the variables of the first
forward DBT as follows:
\be\label{Rbeta1}
\bbeta^{(1)}_n(i) = \frac{b_n(i)}{b_n(i) - \balpha^{(1)}_n(i)}
\ee
The system obtained from (\ref{Rb1}) can be easily reduced to
(\ref{fa1})-(\ref{fz1})
and is not considered here anymore.
\subsection{Continuum limit}

Introducing some discrete shift of time  $\;\Delta\;>\;0\;$, one can rewrite
all the equations describing the first forward DBT as follows:
\be \label{Rf11}
\Phi_n(t+\Delta) = \Phi_n(t) + \balpha^{(1)}_n(t) \Phi_{n-1}(t)
\ee
\be\label{fa11}
a_n(t+\Delta) = a_{n-1}(t)\;\frac{a_n(t)-\balpha^{(1)}_{n+1}(t)}
{a_{n-1}(t)-\balpha^{(1)}_{n}(t)}
\ee
\be\label{fb11}
b_n(t+\Delta) = b_{n-1}(t)\;\frac{b_n(t)-\balpha^{(1)}_{n}(t)}
{b_{n-1}(t)-\balpha^{(1)}_{n-1}(t)}
\ee
In order to get the proper limit in (\ref{fz1}) one should rescale the
constants $\;z_i\;$; for simplicity, we assume now that they do not depend on
$\;i\;$:
\be
z_i \to g\;\Delta
\ee
Then we get
\be\label{fz11}
g\Delta\;\frac{b_n(t)}{\balpha^{(1)}_n(t)} =
g\;\Delta - a_n(t) + \balpha^{(1)}_{n+1}(t)
\ee
and solution to (\ref{fz11}) has the asymptotics
\be
\balpha^{(1)}_n(t) = -g\Delta\;\frac{b_n(t)}{a_n(t)}\;+\;O(\Delta^2)
\ee
and, therefore, from (\ref{Rf11})-(\ref{fb11})
\be
\frac{1}{\Delta}\Bigl\{\Phi_n(z,t+\Delta) - \Phi_n(z,t)\Bigr\} \;=\;
-\;g\;\frac{b_n(t)}{a_n(t)}\Phi_{n-1}(z,t)\;+\;O(\Delta)
\ee
\be
\frac{1}{\Delta}\Bigl\{a_n(t+\Delta) - a_n(t)\Bigr\}\;=\;
-\;g\;\left(\frac{b_{n}(t)}{a_{n-1}(t)}-
\frac{b_{n+1}(t)}{a_{n+1}(t)}\right)\;+\;O(\Delta)
\ee
\be
\frac{1}{\Delta}\Bigl\{b_n(t+\Delta) - b_n(t)\Bigr\}\;=\;
-\;g\;b_n(t)\left(\frac{1}{a_{n-1}(t)}-\frac{1}{a_{n}(t)}\right)\;+\;
O(\Delta)
\ee
and, in the continuum limit, $\,\Delta \to 0\,$ these equations lead
exactly to (\ref{a-eq}), (\ref{b-eq}).

The analogous equations can be written for the second forward DBT
but now $\;z_i\;$ should be rescaled as follows:
\be
z_i \to \frac{1}{g\;\Delta}
\ee
and (\ref{Rf2}), (\ref{fa2'}) - (\ref{fz2'}) now have the form
\be\label{Rf22}
\Phi_n(t+\Delta) = \bigl(1-\balpha^{(2)}_n(t)\bigr)\Phi_n(t) +
z\balpha^{(2)}_n(t)\Phi_{n-1}(t)
\ee
Substituting to (\ref{Rsp}) gives the equations:
\be\label{fa22'}
a_n(t+\Delta)=
a_n(t)\frac{1-\balpha^{(2)}_{n+1}(t)}
{1-\balpha^{(2)}_{n}(t)})
\ee
\be\label{fb22'}
b_{n}(t+\Delta) =
b_{n-1}(t)\frac{\balpha^{(2)}_n(t)}{\balpha^{(2)}_{n-1}(t)}
\ee
\be \label{fz22'}
a_n(t) + b_n(t)\;\frac{1-\balpha^{(2)}_n(t)}
{\balpha^{(2)}_n(t)} =
\frac{1}{g\;\Delta}\;\frac{1}{1-\balpha^{(2)}_{n+1}(t)}
\ee
{}From (\ref{fz22'}), it follows that
\be
\balpha^{(2)}_n(t) \simeq g\Delta\;b_n(t)\Bigl\{ 1 +
g\Delta\;\Bigl(a_n(t) - b_n(t) - b_{n+1}(t)\Bigr)\Bigr\}
\ee
and, therefore,
\be
\frac{1}{\Delta}\Bigl\{\Phi_n(z,t+\Delta) - \Phi_n(z,t)\Bigr\} \;=\;
-\;g\;b_n(t)\Bigl(\Phi_n(z,t) - z\Phi_{n-1}(z,t)\Bigr)\;+\;O(\Delta)
\ee
\be
\frac{1}{\Delta}\Bigl\{a_n(t+\Delta) - a_n(t)\Bigr\}\;=\;
g\;a_n(t)\Bigl(b_{n}(t) - b_{n+1}(t)\Bigr)\;+\;O(\Delta)
\ee
\be
\frac{1}{\Delta}\Bigl\{b_n(t+\Delta) - b_n(t)\Bigr\}\;=\;
g\;b_n(t)\Bigl( a_n(t) - a_{n-1}(t) + b_{n-1}(t) -
b_{n+1}(t)\Bigr)\;+\;O(\Delta)
\ee
It is clear that, in the limit $\,\Delta \to 0\,$, one reproduces
the continuum equations
(\ref{kuz1}), (\ref{kuz2}).

\subsection{Limit to Toda chain}
Here we very briefly outline the theory of the discrete evolution of the usual
Toda
chain (see, for example, \cite{SZh} and references therein). The spectral
problem
for the Toda chain has the form
\be\label{TC}
\lambda\Psi_n(i|\lambda) = \Psi_{n+1}(i|\lambda) -
p_n(i)\Psi_n(i|\lambda)+ R_n(i)\Psi_{n-1}(i|\lambda)
\ee
where we consider the whole set of successive DBT
labelled by the discrete index $\;i\;$  as a discretized variant of the
continuous evolution equations
\be\label{forw}
\delta\Psi_n(i)\equiv \Psi_n(i+1) - \Psi_n(i) = A_n(i) \Psi_{n-1}(i)
\ee
which should be compatible with the spectral problem (\ref{TC}).
(\ref{forw}) describes the forward DBT.
Analogously, the backward DBT is
\be\label{back}
\Psi_n(i|\lambda) = \frac{1}{\lambda - \lambda_i}
\Bigl\{\Psi_{n+1}(i+1|\lambda) +
B_{n}(i)\Psi_n(i+1|\lambda)\Bigr\}
\ee
where coefficients $\;B_n(i)\;$ should be determined. The compatibility
conditions
give the following system of the evolution equations:
\be \label{D1}
p_n(i) = -\lambda_i - A_{n+1}(i) - B_{n}(i)
\ee
\be \label{D2}
p_n(i+1) = -\lambda_i  - A_{n}(i) - B_{n}(i)
\ee
\be \label{D3}
R_n(i) = A_n(i)B_{n}(i)
\ee
\be  \label{D4}
R_n(i+1) = A_n(i)B_{n-1}(i)
\ee
where $\;\lambda_i\;$ serve as free parameters.
In order to get the close connection with the general theory of
integrable systems as well as with the theory of orthogonal polynomials,
one can introduce new variables $\;h_n(i) \equiv \exp\bigl(q_n(i)\bigr)\;$
through the relation:
\be\label{rn}
R_n(i)\; \equiv\; \frac{h_n(i)}{h_{n-1}(i)}
\ee
Then, in the terms of variables $\;h_n(i)\;$
\be \label{An}
A_n(i) = \frac{h_n(i+1)}{h_{n-1}(i)}
\ee
\be \label{Bn}
B_{n}(i) = \frac{h_{n}(i)}{h_{n}(i+1)}
\ee
and, from (\ref{D1})-(\ref{D2}), one can get the following evolution
equations for the discrete Toda chain (assuming for simplicity that
parameters $\,\lambda_i\,$ do no depend on $\,i\,$):
\be\label{udt}
\frac{h_n(i+1)h_n(i-1)}{h_n^2(i)} =
\frac{1-\displaystyle{\frac{h_{n}(i+1)}{h_{n-1}(i-1)}}}
{1-\displaystyle{\frac{h_{n+1}(i+1)}{h_{n}(i-1)}}}
\ee
Now we can consider the transition from the RTC to the usual discrete-time
Toda chain. Let (compare with (\ref{a-rep}), (\ref{b-rep}))
\be\label{Rlim1}
a_n(i) \simeq 1-\ep p_n(i)\;\;;\;\;\;\;
b_n(i) \simeq  - \ep^2R_n(i)\\
z \simeq 1+\ep\lambda\;\; ;\;\;\;\;  z_i \simeq 1+\ep\lambda_i
\ee
Introduce also functions $\Psi_n(i)$\footnote{Let us note that, under this
limit, the corresponding biorthogonal polynomials turn into the usual
Toda orthogonal polynomials, with the biorthogonality relation
(\ref{orth-cond})
transforming to the orthogonality relation. This is why we sometimes call these
biorthogonal polynomials relativistic ones.}
\be\label{Rlim2}
\Phi_n(i) \simeq \ep^n\Psi_n(i)
\ee
It is easy to see that (\ref{Rf1}) leads to the forward DBT for the Toda chain
if one identifies
\be
\balpha^{(1)}_n(i) \simeq \ep A_n(i)
\ee
In this case, for example, equation (\ref{fz1}) reduces to (\ref{D1}).
Actually, it is easy to see that the whole system (\ref{fa1})-(\ref{fz1})
reduces to (\ref{D1})-(\ref{D4}) in the limit $\,\ep \to 0\,$. Moreover,
from (\ref{Rbeta1}) one gets
\be
\bbeta^{(1)}_n(i) = \ep\frac{R_n(i)}{A_n(i)} + O(\ep^2)\; \simeq \ep B_n(i)
\ee
thus obtaining the proper limit of the first backward DBT of the RTC
(\ref{Rb1})
to that of the Toda chain (\ref{back}).

It is clear also that, in this limit, the second forward DBT (\ref{Rf2})
reduces to (\ref{forw}) and all the equations we get from (\ref{fa2'})-
(\ref{fz2'}) are equivalent to (\ref{D1})-(\ref{D4}) when $\, \ep \to 0\,$.

There exist some other interesting limits. In particular, one can consider
the degenerate case of evolution equations (\ref{fa1})-(\ref{fz1})
in the limit $\,z_i \to \infty$. In this case, the solution to (\ref{fz1})
has the asymptotics
\be
\balpha^{(1)}_n(i) = b_n(i) +
\frac{1}{z_i}b_n(i)\Bigl(a_n(i)- b_{n+1}(i)\Bigr) +
O\Bigl(\;\frac{1}{z_i^2}\;\Bigr)
\ee
Thus, in this limit equations (\ref{fa1}) and (\ref{fb1}) take the simple form:
\be\label{dis4}
a_n(i+1) = a_{n-1}(i)\;\frac{a_n(i)-b_{n+1}(i)}{a_{n-1}(i)-b_{n}(i)}
\ee
\be\label{dis5}
b_n(i+1) = b_{n}(i)\;\frac{b_{n+1}(i)-a_{n}(i)}
{b_{n}(i)-a_{n-1}(i)}
\ee
We should stress that these equations can be easily derived from
the spectral problem (\ref{Rsp}) just representing it in the form
\be\label{}
\Phi_{n+1}(i|z) +a_n(i)\Phi_n(i|z) = z\;\Phi_n(i+1|z)
\ee
\be
\Phi_n(i+1|z)=\Phi_n(i|z) + b_n(i)\Phi_{n-1}(i|z)
\ee
Using representation (\ref{a-rep}), (\ref{b-rep}) we can rewrite system
(\ref{dis4}), (\ref{dis5}) in terms of coordinates $\,q_n(i)\,$ as follows:
\be\label{sdt}
\exp\bigl\{q_{n}(i+1) - 2q_{n}(i) + q_{n}(i-1)\bigr\} =
\frac{1+\epsilon^2\exp\bigl\{q_{n+1}(i) - q_{n}(i)\bigr\}}
{1 +\epsilon^2\exp\bigl\{q_{n}(i) - q_{n-1}(i)\bigr\}}
\ee
thus getting the counterpart of the discrete-time Toda chain \cite{Sur}.
We should stress,
however, that equations (\ref{udt}) and (\ref{sdt}) are obtained as different
limits from the general discrete RTC (\ref{fa1})-(\ref{fz1}). Moreover,
they are described by different Lax operators.

\sect{Some applications to biorthogonal polynomials}
In this section we discuss some peculiar properties of the system of
(relativistic) polynomials leading to the RTC, biorthogonal with some specific
measures.

\subsection{Finite systems of orthogonal polynomials}

We already discussed in sect.3 that the relativistic polynomials
satisfy recurrent relations (\ref{u-rec})
\be
\Phi_{n+1}(z) = z\Phi_{n}(z)+ S_{n}z^{n} \Phi^{\star}_{n}(z^{-1})\\
\Phi^{\star}_{n+1}(z^{-1}) = z^{-1}\Phi^{\star}_{n}(z^{-1})+
S^{\star}_{n}z^{-n}\Phi_{n}(z)
\ee
provided
the polynomials are normalized to be the
n-th order monic (i.e. with the coefficient in front of the leading term
equal to 1) polynomials. In particular, the initial conditions are
$\Phi_0(z)=\Phi^{\star}_0(z)=1$. We also demonstrated that such polynomials
satisfy the RTC recurrent relations (\ref{rec2})
\be
\Phi_{n+1}(z) + a_{n}\Phi_{n}(z) =
z \{\Phi_{n}(z) + b_{n}\Phi_{n-1}(z)\}\;\; ,\;\;\;\;\;n\in\ZZ
\ee
As far as we know, it was G.Baxter who first
considered the system (\re{u-rec})  \ci{Baxter} (for recent
development of the theory of these polynomials see e.g.
\ci{Pastro,AlIs,IsMas}).  On the other hand, Pastro showed
\ci{Pastro} that the theory of the polynomials defined by (\re{u-rec})
is closely
related to the Laurent orthogonal polynomials first introduced in
the framework of the strong Stiltjes moment problem \ci{JTW}. It is interesting
to
note that the RTC was studied in \ci{comm} by the method of continued
T-fractions arising in the context of the same strong Stiltjes moment problem,
however, without using the orthogonal polynomials.

Note that in general, $S_n$ and $S^{\star}_n$  are arbitrary complex
parameters. When $S_n=\bar S^{\star}_n$ and $|S_n|<1$, one gets the
theory of Szeg\"o polynomials orthogonal on the unit circle \ci{sz,ger}.
The case of $|S_n|>1$ is much less trivial and has not been considered in
detail.

It is interesting to discuss how given
recurrent relations ((\re{u-rec}) or (\re{rec2})) are connected to the concrete
orthogonality measure. In general, when the space of
the polynomials is infinite-dimensional, one can hardly formulate any
statement on the existence and support of the corresponding measure. However,
for the finite-dimensional representations corresponding to the Toda molecule
such (discrete) measures can be found explicitly.

For doing this, note that one can easily derive from (\re{u-rec})
the identity
\be
y^{-n}({\tilde \Phi}^{\star}_n(y)\Phi_{n+1}(x)-
{\tilde \Phi}^{\star}_n(x)\Phi_{n+1}(y))h_n^{-1}=
(x-y) \sum_{k=0}^{n}{\Phi_k(x)\Phi^{\star}_k(y^{-1})h_k^{-1}}
\lab{dc} \ee
where $h_0=1$
and we introduced for the sake of brevity ${\tilde \Phi}^{\star}_n(z)\equiv
z^n\Phi^{\star}_n(1/z)$.
This identity is valid for arbitrary complex $x$ and $y$  unless $x\ne y$. For
$x=y$ we have the identity
\be
(y^n h_n)^{-1}({\tilde \Phi}^{\star}_n(y)\Phi_{n+1}^\prime (y)-
\Phi_{n+1}(y){\tilde \Phi}_n^{\star\prime} (y))=
\sum_{k=0}^{n}{\Phi_k(y) \Phi^{\star}_k(1/y)h_k^{-1}}
\lab{con} \ee
Identities (\re{dc}) and (\re{con})
can be considered as "relativistic" analogs
of the well known Darboux-Christoffel formula in the theory of the ordinary
orthogonal polynomials \ci{sz}.

Now let us turn to the finite set of polynomials. In order to reduce our
system,
we put $h_{n+1}=0, \quad h_n \ne 0$ (this means that $S_n S^{\star}_n
=1$ and this corresponds to the Toda molecule case of section 6).
Let $y_j,\quad j=0,1,\dots,n$ be the roots of the polynomial
$\Phi_{n+1}(y)$. We assume additionally that all these roots are simple. Then,
from (\re{dc}) and (\re{con}), one gets the following orthogonality relation
\be
\sum_{k=0}^n{\Phi_k(y_s)\Phi^{\star}_k(y_r^{-1})h_k^{-1}}= w_s^{-1} \delta_{rs}
\lab{ort1} \ee
where
\be
w_s=\frac{h_n}{\Phi^{\star}_n(y_s^{-1})\Phi_{n+1}^\prime(y_s)}
\lab{w} \ee

At last, using that, for the orthogonal matrices, their transponed are
orthogonal too, one finally obtains the dual orthogonality
relation
\be \sum_{k=0}^n {w_k \Phi_n(y_k) \Phi^{\star}_m(y_k^{-1})}=h_n \delta_{nm}
\lab{ort2} \ee
Therefore, $w_k$ is the discrete weight function whose support is the set of
zeros of the polynomial $\Phi_{n+1}(z)$.

Note that because of the condition $S_n S^{\star}_n =1$, the roots $y_k$
of the polynomial $\Phi_{n+1}(z)$ coincide with the roots of the polynomial
$\Phi^{\star}_{n+1}(1/z)$. Then, using
the symmetry between the polynomials $\Phi_n(z)$ and
$\Phi^{\star}_n(z)$, one derives another orthogonality relation
\be
\sum_{k=0}^n \tilde w_k \Phi_n(y_k^{-1}) \Phi^{\star}_m(y_k)=h_n \delta_{nm}
\lab{ort3} \ee
where
\be
\tilde w_k = \frac{h_n}{\Phi_n(y_k)\Phi_{n+1}^{\star\prime}(y_k^{-1})}
\lab{w1p} \ee

In general, the weight functions $w_k$ and $\tilde w_k$ do not coincide and,
therefore,
we have two different orthogonality relations for the same polynomials.
However, if $S_k=S^{\star}_k,\quad k=0,1,\dots,n$ (i.e. we deal with the
symmetric case) then $w_k=\tilde w_k$. Moreover, in this case the polynomials
$\Phi_k(z)$ and $\Phi^{\star}_k(z)$ coincide and, hence, the set of the
roots {$y_k$} of the polynomial $\Phi_{n+1}$ coincides with the set
{$y_k^{-1}$}
of roots of
the polynomial $\Phi^{\star}_{n+1}(z)$. This means that, in the symmetric case,
the
polynomial $\Phi_{n+1}(z)$ is invertible, i.e. the coefficients in
front of the terms $z^j$
and $z^{n+1-j}$ coincide (or, equivalently, all the roots enter with the
pairs $y_k,\quad y_k^{-1}$). In particular, when $|S_k|<1$, all the
roots lie on the unit circle.

\subsection{Anzatz of separated variables}

Now let us consider another interesting peculiar case of the relativistic
polynomials connected with a specific solution to the RTC hierarchy. Namely,
look at
the following ansatz of separated  variables (see (\ref{rec2}))
(for the non-relativistic Toda chain
this ansatz was studied in \ci{zhe})
\be a_n(t)=\gamma_1 (t) \nu_n, \quad
b_n(t)=\gamma_2 (t) \mu_n \lab{sep} \ee
where $\gamma_i (t)$ depend on $t$
only whereas $\nu_n$ and $\mu_n$ depend only on $n$. Putting $b_0=0$
(because of (\re{polyn})), one gets the solution
\be
a_n(t)=-\frac{t}{n+c+1}, \quad b_n(t)=-\frac{tn}{(n+c)(n+c+1)}
\lab{sol} \ee
where $a$ is an arbitrary constant. Then, the
corresponding coefficients $S_n$ and
$S^{\star}_n$ are equal to
\be \alpha_n(t)=-\frac{t^n}{(c+1)_{n+1}},\quad
\beta_n(t)=-t^{-n}(c)_{n+1}
\lab{ab} \ee where $(c)_n=c(c+1)\dots (c+n-1)$ is
the Pochhammer symbol.

In order to find explicit expressions for the polynomials $\Phi_n(z)$ and
$\Phi^{\star}_n(z)$, note that the recurrent coefficients (\re{ab}) or
(\re{sol})
can be obtained
as some limiting case of the system of biorthogonal
polynomials on the unit circle proposed by R.Askey \ci{as}. Indeed, the Askey
polynomials have the recurrent coefficients
\be
S_n= -\frac{(b)_{n+1}}{(c+1)_{n+1}}, \quad
S^{\star}_n=-\frac{(c)_{n+1}}{(b+1)_{n+1}}
\lab{asc} \ee
where $a,b$ are arbitrary parameters. Corresponding polynomials are expressed
in terms of the Gauss hypergeometric function \ci{as}
\be
\Phi_n(z)=\frac{(b)_n}{(c+1)_n}\: {_2}F{_1} \left({-n,1+c \atop 1-n-b};z
\right),
\quad
\Phi^{\star}_n(z)= \frac{(c)_n}{(b+1)_n}\: {_2}F{_1} \left({-n,1+b \atop
1-n-c};z
\right)
\lab{asp} \ee

These polynomials are biorthogonal on the unit circle
\be
\int_{-\pi}^{\pi} \Phi_n(e^{i\theta})
\Phi^{\star}_m(e^{-i\theta}) e^{-i\theta (c-b)/2}
|\sin(\theta /2)|^{c+b} d\theta = 0 , \quad n\ne m
\lab{asw} \ee

It is easy to see that the recurrent coefficients (\re{ab}) can be obtained
from
(\re{asc})
by the trivial rescaling in the limit $b\to \infty$. Omitting the simple
technical details, we present here only the result.
The polynomials corresponding to
solution (\re{sol}) or (\re{ab}) are
\be
\Phi_n(z;t)=\frac{t^n}{(c+1)_n}\: {_2}F{_0} (-n,1+c;-z/t), \quad
\Phi^{\star}_n(z;t)=t^{-n}(c)_n \: {_1}F{_1}\left({-n \atop 1-n-c};tz \right)
\lab{solt} \ee
These polynomials are biorthogonal on the unit circle
\be
\int_{-\pi}^{\pi}{\Phi_n(e^{i\theta})\Phi^{\star}_m(e^{-i\theta}) e^{ic\theta}
\exp(-te^{-i\theta})}=0, \quad n\ne m.
\lab{biort2} \ee

Now let us consider the first Darboux-B\"acklund transformation
(\ref{dis4})-(\ref{dis5}) (see the previous section).
We define
$w(z)$ to be a weight function for the biorthogonal polynomials
\be
\int{\Phi_n(z)\Phi^{\star}_m(z^{-1})w(z)dz}=0, \quad n \ne m
\lab{defwz} \ee
Then, it is transformed under the first Darboux-B\"acklund transformation as
\be
w(z;\tau+1)=\kappa zw(z;\tau) \lab{discrw} \ee
where $\kappa$ is a normalization constant.

In particular, in the case of the separate variables
solution (\re{sol}), this
transformation is equivalent to shifting the parameter
$c$: $\tilde c =c+1$, and, thus, one has a whole family of solutions depending
on
the both continuous and discrete times
\be
b_n(t;\tau)=-\frac{tn}{(n+c_0+\tau)(n+c_0+\tau+1)}, \quad
a_n(t;\tau)=-\frac{t}{n+c_0+\tau+1}
\lab{full} \ee
The property (\re{discrw}) is obviously fulfilled, which can be seen from
orthogonality relation (\re{biort2}).

It is interesting to note that the Askey polynomials (\re{asp})
themselves obey
the discrete time Toda dynamics determined by the Darboux-B\"aklund
transformation provided that
$c(\tau)=c_0 +\tau, \quad b(\tau)=b_0-\tau$.

The general Darboux-B\"aklund transformations of the biorthogonal polynomials
will be considered in a separate publication.

\section{Concluding remarks}
In the present paper we considered some different representations of the
relativistic
Toda hierarchy, which are naively non-related to each other. However, from the
point
of view of studying the RTC hierarchy itself, the most promising representation
is
that describing the relativistic Toda hierarchy as a particular
reduction
of the two-dimensional Toda lattice hierarchy. However, even this quite large
enveloping hierarchy is still insufficient.

Indeed, the RTC-reduction of the 2DTL being transparent on the
level
of the Lax operators can not be explicitly written in terms of the Clifford
element. In other words, no reasonable description of the point of the
Grassmannian is
possible in the context of the usual (one-component) Toda lattice. Loosely
speaking,
the Toda lattice is too "rigid" to reproduce both the continuous and discrete
flows of the
 RTC.
Therefore, one can try to embed the RTC in a more general system which admits
more natural
reductions.

This is done in the forthcoming publication \cite{KMZ}, where we
show that the RTC has a nice interpretation if considering it
as a simple reduction of the two-component KP (Toda) hierarchy. In this case,
any
dynamical variables acquire two discrete indices (we have two fermionic vacua
in this picture) so that one can treat the second
vacuum number as a discrete time which
generates the sequence of the degenerate  B\"acklund transformations
(or, equivalently, the discrete evolution of the modified Toda-type).
On the other hand, we need yet another discrete index in order to generate the
whole
discrete-time RTC. It turns out that the natural interpretation of this
additional
discretization is possible if one introduces the so-called Miwa variables (with
the corresponding multiplicities) \cite{DJKM}
after imposing the reduction conditions. The whole evolution of the RTC can be
interpreted
now as follows: the evolution along the additional vacuum generates the
modified Toda
equations while the evolution with respect to the multiplicities of the Miwa
variables
leads to  general Darboux-B\"acklund transformations. This is the reason why
the (properly) reduced two-component KP (Toda) is the true framework for
description of
RTC. Moreover, it turns out that, in the framework
of the 2-component hierarchy, the continuous AKNS system, Toda chain hierarchy
{\it and}
the discrete AKNS (which is equivalent to the RTC how we proved in this paper)
can be treated on equal footing. In particular, this means that all these
systems
corresponds to the same subspace in the Grassmannian.

\section*{Acknowledgments}
We acknowledge V.Fock, A.Marshakov and A.Zabrodin for the discussions.
A.Z. is also grateful to V.Spiridonov, S.Suslov and L.Vinet for stimulating
discussions. S.K. and A.M. express their gratitude to O.Lechtenfeld and
S.Ketov for the kind hospitality in the Hannover University. The work of
S.K. is partially supported by grants RFFI-95-02-03379, ISF-MGK000
and by  Volkswagen Stiftung, that of A.M. -- by grants RFFI-95-01-01106,
ISF-MGK000, INTAS-93-1038 and by Volkswagen Stiftung, and the work of A.Z --
by ISF grant U9E000.

\app{Evolution equations from the orthogonality conditions}

Here we shall give some examples of deriving the evolution equations
in the polynomial case starting from the orthogonality condition (\ref{sp}),
(\ref{orth-cond}). In order to keep the close connection with the
general theory of the monic orthogonal polynomials, we write down here the
spectral problem directly in terms of $\Phi_{n}$ and $\Phi_n^{\star}$:
\be \label{aq1}
z\Phi_{n}(z) = \Phi_{n+1}(z)
    - S_{n}h_{n}\sum_{k=0}^{n}\frac{S^{\star}_{k-1}}{h_{k}}\Phi_{k}(z)
                \equiv {\cal L}_{nk}\Phi_{k}(z)
\ee
\be \label{aq2}
z^{-1}\Phi^{\star}_{n}(z^{-1}) = \Phi^{\star}_{n+1}(z^{-1})
    - S^{\star}_{n}h_{n}\sum_{k=0}^{n}
         \frac{S_{k-1}}{h_{k}}\Phi^{\star}_{k}(z^{-1})
    \equiv h_{n}\ov{\cal L}_{kn}\frac{1}{h_{k}}
\Phi^{\star}_{k}(z^{-1})
\ee
Now the differentiation of (\ref{orth-cond}) with respect to $t_{-1}$
gives (for $n>k$) :
\be
\left<\frac{\d \Phi_{n}}{\d t_{-1}},\Phi^{\star}_{k}\right>
= <\Phi_{n},z^{-1}\Phi^{\star}_{k}>
\ee
Using (\ref{aq2})  it is easy to see that
\be \label{t-1}
\frac{\d \Phi_{n}}{\d t_{-1}} =
\frac{h_{n}}{h_{n-1}} \Phi_{n-1}  \equiv
(1-S_{n-1}S^{\star}_{n-1})\Phi_{n-1} \equiv
\frac{a_n}{b_n}\Phi_{n-1}
\ee
Similarly, the differentiation of (\ref{orth-cond}) with respect to
$t_{1}$ gives (for $n>k$) due to (\ref{aq1})
\be
\left<\frac{\d \Phi_{n}}{\d t_{1}},\Phi^{\star}_{k}\right>
= -<z\Phi_{n},\Phi^{\star}_{k}> = S_{n}S^{\star}_{k-1}h_{n}
\ee
Therefore,
\be \label{t-2}
\frac{\d \Phi_{n}}{\d t_{1}}
= S_{n}h_{n}\sum_{k=0}^{n-1}\frac{S^{\star}_{k-1}}{h_{k}}\Phi_{k}(z)=\\
=  \frac{S_{n}}{S_{n-1}}\frac{h_{n}}{h_{n-1}}(\Phi_{n} -z \Phi_{n-1})
\equiv\; -\;b_n(\Phi_{n} -z \Phi_{n-1})
\ee
The equations (\ref{t-1}), (\ref{t-2}) are nothing but the simplest
evolution equations (\ref{todaev}), (\ref{todaev2}) which determine
the RTC. Note that all the general equations (\ref{OE1}) can be obtained
in the same way. For example, the differentiation of (\ref{orth-cond})
for $\;n>k\;$ with respect to $\;t_m\;$ gives:
\be
\left<\frac{\d \Phi_{n}}{\d t_{m}},\Phi^{\star}_{k}\right>
= - <({\cal L}^m)_{ns}\Phi_{s},\Phi^{\star}_{k}> =
- ({\cal L}^m)_{nk}h_k\theta(n-k-1)
\ee
and, therefore,
\be
{\partial \Phi_n\over \partial t_m} =
- [({\cal L}^m)_-]_{nk}\Phi_k
\ee
Similarly, the differentiation of (\ref{orth-cond}) for $\;k=n\;$
yields
\be
\frac{\d h_{n}}{\d t_{m}}\; =\; <z^m\Phi_n,\Phi^{\star}_n>\; \equiv\;
<({\cal L}^m)_{nk}\Phi_k,\Phi^{\star}_n>\; = \;({\cal L}^m)_{nn}h_n
\ee
In the particular case of $\;m=1\;$, one gets just $\;{\cal L}_{nn} =
-S_nS^{\star}_{n-1}\;$, which leads to (\ref{ht1}) with $\;\gamma = 0\;$.

\newpage
\app{Lax operators in 2DTL framework}

In this appendix we are going to represent the spectral problem (\ref{q1})
arising from the orthogonal polynomials
as the "usual" spectral problem.
First of all, in the fast-decreasing case
\be
S_n  \to 0 \;\; , \;\;\;\;\;\; S^{\star}_n  \to 0 \;\;\;\; ;
\;\;\; n\to \pm \infty
\ee
one should prove that the solution to (\ref{q1}) with asymptotics
(\ref{p1}) satisfies the spectral problem \footnote{See identification
(\ref{PP})} \be \label{lq1} {\cal L}_{nk}{\cal P}_{k}(z) =
z {\cal P}_{n}(z)
\ee
where ${\cal L}_{nk}$ is given by (\ref{RTC1}). Indeed, the equation
(\ref{q1}) can be re-written in the form
\be\label{qq}
\frac{1}{S_{k}h_{k}}\Bigl({\cal P}_{k+1}(z) - z{\cal P}_{k}(z)\Bigr) -
\frac{1}{S_{k-1}h_{k-1}}\Bigl({\cal P}_{k}(z) - z{\cal P}_{k-1}(z)\Bigr) =
\frac{S^{\star}_{k-1}}{h_{k}}{\cal P}_{k}(z)
\ee
In order to get (\ref{lq1}), one should sum (\ref{qq})
from $k=-\infty$ to $k=n$ provided the following boundary condition
is fulfilled
(using that \quad $h_k \to Const$ \quad when $k \to -\infty$) :
\be \label{zeta}
\zeta_{k}(z) \equiv
\frac{1}{S_{k}}\Bigl({\cal P}_{k+1}(z) - z{\cal P}_{k}(z)\Bigr) \to 0\;\; ;
\;\;\;\;\;k \to -\infty
\ee
With the help of (\ref{w-p1}), (\ref{w1}), it is easy to see that
coefficients
$w_{i}(k)$ satisfy the recurrent relations
\be
\frac{1}{S_{k}}\Bigl\{w_{i+1}(k+1) - w_{i+1}(k)\Bigr\} -
\frac{1}{S_{k-1}}\Bigl\{w_{i}(k) - w_{i}(k-1)\Bigr\} =
S^{\star}_{k-1}w_{i}(k-1)\;\;\; ;\;\;\;\;\; w_{0}(k) \equiv 1
\ee
i.e
\be \label{w-rec}
\frac{1}{S_{k}}\Bigl\{w_{i}(k+1) - w_{i}(k)\Bigr\}
= \sum_{j=1}^{i}S^{\star}_{k-j}w_{i-j}(k-j)
\ee
Thus, for any {\it fixed} $\;i\;$, the r.h.s of (\ref{w-rec}) tends
to zero provided $\;S^{\star}_{k}w_{i}(k) \to 0$ as $\;k \to -\infty$.
For sufficiently fast-decreasing $\;S_k$, $S^{\star}_k\;$, it is really true.
Therefore, we see that all the coefficients of $\;\zeta_k(z)\;$ in
(\ref{zeta}) vanish in the limit $\;k \to -\infty$. Since solution
${\cal P}_n^{(1)}(z)$ is defined in the range of {\it large} values
of the spectral parameter $\;z\;$, the whole function $\;\zeta_{k}(z)\;$
is convergent in the limit of large negative $\;k\;$ and, thus,
satisfies (\ref{zeta}). Note that the same arguments fail in the case
of solution (\ref{p2}) since it is defined at {\it small}\quad values
of the spectral parameter and the corresponding function $\;\bar\zeta_{k}(z)\;$
can not be properly defined as $k \to -\infty$ and, therefore,
${\cal P}^{(2)}_n(z)$ does not satisfy (\ref{lq1}) in general.

To understand this situation in more details, it is instructive to
consider the case of the forced hierarchy when
\be\label{forcedS}
S_{k} = S^{\star}_{k} \equiv 1\;\; ; \;\;\; k<0\\
S_{k}\;,\; S^{\star}_{k} \to 0\;\; ; \;\;\;\; k \to +\infty
\ee
The first condition means also (see (\ref{hS-rel})) that
\be
\frac{h_{k+1}}{h_{k}} = 0 \;\;\;\; k<0\;\; \; h_0 \neq 0
\ee
In the forced case,  ${\cal P}_n^{(1)}(z)$ can be easily
described. Indeed, it is a polynomial solution to (\ref{q1}) of the type
(\ref{polyn}) at $\;\;\;n > 0\;$ , ${\cal P}_0(z) =1$ and
\be
{\cal P}_{n}^{(1)}(z)  = (z + 1)^{n}\;\; , \;\;\;\;\; n<0
\ee
and it is well defined in the whole region of the spectral parameter.
This solution is non-singular in the sense that
\be\label{non-sing}
\frac{1}{h_n}{\cal P}_{n}^{(1)}(z) = 0\;\;,\;\;\;\; n<0
\ee
Thus, from (\ref{qq}) in the forced case, one can get the same spectral
equation as in the fast-decreasing case, i.e
\be \label{PPP}
z{\cal P}_n(z) = {\cal P}_{n+1}(z) -
S_nh_n\sum ^n_{k=-\infty} {S^{\star}_{k-1}\over h_k}
{\cal P}_k(z) \equiv  {\cal L}_{nk}{\cal P}_k(z)
\ee
where the summation actually goes from $0\;$ to $\;n$ due to
(\ref{non-sing}).

On the other hand, the second independent solution to (\ref{q1}),
$\;{\cal P}^{(2)}_n(z)\;$, is non-polynomial even in the forced case
and can be treated as a formal series at small values of $\;z\;$ only.
Due to asymptotics (\ref{p2}), it is singular, i.e.
\be\label{sing}
\frac{1}{h_n}{\cal P}_{n}^{(2)}(z) \neq 0\;\;,\;\;\;\; n<0
\ee
Moreover, putting
\be \label{p2-x}
{\cal P}^{(2)}_n(z) = \frac{1}{h_{n}}x_n(z)
\ee
one can find out the exact answer
\be\label{x}
x_n(z) = z^n(1+z)^{-n-1}\;\;, \;\;\;\;\;n<0
\ee
Now it is easy to see the reason why ${\cal P}^{(2)}_n(z)$ does not satisfy
(\ref{PPP}) in the generic situation. This is since the series
\be
\sum_{k=-\infty}^{-1}{S^{\star}_{k-1}\over h_k}{\cal P}^{(2)}_k(z) =
\frac{1}{z}\sum_{k=0}^{\infty}\left(1+\frac{1}{z}\right)^{k}
\ee
is divergent at small $\;z\;$ though {\it formally}\quad it is equal to
$\;-1\;$ and this formal answer miraculously coincides with the
{\it exact}\quad expression which can be obtained for ${\cal P}^{(2)}_n(z)$
from (\ref{qq}) after summation from $k=0$ to $k=n$ and using (\ref{p2-x}),
(\ref{x}). This is a typical example of how the formal analysis leads to a
wrong conclusion and the analytical consideration is of great importance.

Instead of doing all this, one can prove in the same way that
${\cal P}^{(2)}_n(z) \equiv \ov{w}_n(z)$ (see (\ref{w-p2}), (\ref{w2}))
satisfies
\be \label{lq2}
\ov{\cal L}_{nk}{\cal P}_{k}(z) = z^{-1} {\cal P}_{n}(z)
\ee
where $\ov{\cal L}_{nk}$ is given by (\ref{RTC2}). It can be done by
representing (\ref{q1}) in the form
\be
\frac{1}{S_k}\Bigl(z^{-1}{\cal P}_{k+1}(z) - {\cal P}_k(z)\Bigr) -
\frac{1}{S_{k-1}}\Bigl(z^{-1}{\cal P}_{k}(z) - {\cal P}_{k-1}(z)\Bigr) =
S^{\star}_{k-1}{\cal P}_{k-1}(z)
\ee
and performing the summation from $\;k=n\;$ to $\;k=+\infty \;$. The only
thing we need to prove is that ${\cal P}^{(2)}_k(z)$ satisfies the
boundary condition
\be \label{bar-zeta}
\bar \zeta_{k}(z) \equiv
\frac{1}{S_{k}}\Bigl(z^{-1}{\cal P}_{k+1}(z) - {\cal P}_{k}(z)\Bigr)
\to 0\;\; ; \;\;\;\; k \to +\infty
\ee
in the region of small $\;z\;$. This is in complete analogy with the
proof of (\ref{zeta}).

The rest of equations
in (\ref{laxTL}), (\ref{adLax}) can be derived from (\ref{P-bar}) using
the same methods.

\end{document}